\documentclass[12pt]{article}
\pdfoutput=1
\usepackage{subfigure}
\usepackage{amssymb,amsmath}
\usepackage{graphicx}
\usepackage{color}
\usepackage[colorlinks=true
,urlcolor=blue
,citecolor=blue
,linkcolor=blue
,pagecolor=blue
,linktocpage=true
,pdfproducer=medialab
]{hyperref}
\usepackage[a4paper]{geometry}

\usepackage{placeins}
\usepackage[utf8]{inputenc}

\makeatletter \renewcommand{\@dotsep}{10000} \makeatother

\mathchardef\mhyphen="2D

\newcommand{\beq}{\begin{equation}}
\newcommand{\eeq}{\end{equation}}
\newcommand{\bea}{\begin{eqnarray}}
\newcommand{\eea}{\end{eqnarray}}

\newcommand{\mgut}{M_{{\rm GUT}}}
\newcommand{\PS}{SU(4)_{C}\times SU(2)_{L}\times SU(2)_{R}}

\usepackage[nodisplayskipstretch]{setspace}

\begin{document}

\begin{titlepage}
\pagestyle{empty}

\vspace*{0.2in}
\begin{center}
{\Large \bf  Muon $g-2$ in an Alternative Quasi-Yukawa Unification with Low Fine-Tuned Inverse SeeSaw Mechanism
  }\\  
\vspace{1cm}
{\bf  Zafer Alt\i n$^{a,}$\footnote{E-mail: 501407009@ogr.uludag.edu.tr}, \"{O}zer \"{O}zdal $^{b,}$\footnote{E-mail: ozer.ozdal@concordia.ca} and
Cem Salih \"{U}n$^{a,}$\footnote{E-mail: cemsalihun@uludag.edu.tr}}
\vspace{0.5cm}

{\it $^a$Department of Physics, Uluda\~{g} University, TR16059 Bursa, Turkey \\
{\it $^b$Department of Physics, Concordia University 7141 Sherbrooke St. West, Montreal, Quebec, Canada H4B 1R6}
}

\end{center}

\vspace{0.5cm}
\begin{abstract}
\noindent
We explore the low scale implications of the Pati-Salam Model including the TeV scale right-handed neutrinos interacting and mixing with the MSSM fields through the inverse seesaw (IS) mechanism in the light of the muon anomalous magnetic moment (muon $g-2$) resolution, and highlight the solutions which are compatible with the Quasi-Yukawa Unification condition (QYU). We find that the presence of the right-handed neutrinos causes heavy smuons as $m_{\tilde{\mu}} \gtrsim 800$ GeV in order to avoid tachyonic staus at the low scale. On the other hand, the sneutrinos can be as light as about 100 GeV along with the light charginos of mass $\lesssim 400$ GeV, they can yield so large contributions to muon $g-2$ that the discrepancy between the experiment and the theory can be resolved. In addition, the model predicts relatively light Higgsinos ($\mu \lesssim 700$ GeV); and hence the second chargino mass is also light enough ($\lesssim 700$ GeV) to contribute to muon $g-2$. Light Higgsinos also yield less fine-tuning at the electroweak scale, and the regions compatible with muon $g-2$ restricts $\Delta_{EW}\lesssim 100$ strictly, and this region also satisfies the QYU condition. In addition, the ratios among the Yukawa couplings should be $1.8 \lesssim y_{t}/y_{b} \lesssim 2.6$, $y_{\tau}/y_{b}\sim 1.3 $ to yield correct fermion masses. Even though the right-handed neutrino Yukawa coupling can be varied freely, the solutions bound its range to $0.8\lesssim y_{\nu}/y_{b} \lesssim 1.7$. 

\end{abstract}

\end{titlepage}



\section{Introduction}
\label{sec:Intro}

Supersymmetry (SUSY) is one of the forefront candidates for physics beyond the Standard Model (SM). Resolving the gauge hierarchy problem the Higgs boson mass is not too much sensitive to the ultraviolet scale. In addition, minimal supersymmetric version of the SM (MSSM) nicely unifies the three gauge couplings of the SM, and hence, one can identify the unification scale as $\mgut \sim 2\times 10^{16}$ GeV. In this context, SUSY models can study the high energy origins of physics, which is strictly be tested at the low scale experiments by connecting $\mgut$ to the electroweak scale through the renormalization group equations (RGEs). RGEs allow one to build high scale models, and these models can significantly reduce the number of free parameters in comparison to the low scale MSSM models with free parameters more than a hundred. In this approach, minimal SUSY models have been built such as constrained MSSM (CMSSM) and non-universal Higgs models (NUHM), and their phenomenological implications have been excessively explored. These minimal models have been built with the inspiration from $SO(10)$ grand unified theories (GUTs). These GUT models do not only unify the gauge couplings, but the matter fields are also unified into a single representation, since the spinor representation is $16-$dimensional. All the matter fields of a family in MSSM can be resided into such a large representation. In addition, there is still one more space left out, which can be filled naturally by the right-handed neutrinos. In this sense, $SO(10)$ GUTs provide a natural framework to implement the SeeSaw mechanisms through which the neutrinos mix each other and receive non-zero masses favored by the current experiments \cite{Wendell:2010md}.

In addition to the unifications of the gauge couplings and the matter fields, another interesting feature in the GUT models based on the $SO(10)$ gauge symmetry imposed at $\mgut$ is the Yukawa coupling unification (YU) \cite{big-422}. In addition to $SO(10)$ GUT, also the high scale models with Pati-Salam gauge group ($G_{{\rm PS}}=\PS$, hereafter $4-2-2$ for short) \cite{pati-salam}, preserves YU, since it is the maximal subgroup of $SO(10)$. Even though it does not provide a GUT model ($g_{4}\neq g_{L}\neq g_{R}$ in principle), if it breaks into the MSSM gauge group at a scale near by $\mgut$, the gauge couplings receive negligible threshold corrections; and hence, the gauge coupling unification can be maintained in $4-2-2$ as well. In addition,  imposing left-right (LR) symmetry requires $g_{L}=g_{R}\equiv g_{2}$, and consequently $M_{L}=M_{R}\equiv M_{2}$. Even though the hypercharge is not a direct symmetry in $4-2-2$, the hypercharge jenerator can be expressed as 

\begin{equation}
Y=\sqrt{\frac{3}{5}}I_{3R}+\sqrt{\frac{2}{5}}(B-L)
\end{equation}
where $I_{3R}$ and $B-L$ are diagonal generators of $SU(2)_{R}$ and $SU(4)_{C}$ symmetry groups respectively. This relations for the hypercharge generator also yields non-universal gaugino mass terms for the MSSM gaugino fields as

\begin{equation}
M_{1}=\frac{3}{5}M_{2}+\frac{2}{5}M_{3}
\label{422gauginos}
\end{equation}
with $M_{1}$, $M_{2}$, and $M_{3}$ being soft supersymmetry breaking (SSB) mass terms for the MSSM gauginos associated with the $U(1)_{Y}$, $SU(2)_{L}$, and $SU(3)_{C}$ gauge groups respectively.

YU shared with the SUSY high scale models mentioned above provides an exclusive impact on the low scale phenomenology, even though YU is imposed at $\mgut$. This impact is mostly based on the fact that the bottom Yukawa coupling needs to receive the largest negative threshold corrections in order to yield consistent masses for the top and bottom quarks \cite{Gogoladze:2010fu}. Even though it is a very effective condition from $\mgut$ to shape the low scale parameter space, YU fails to yield consistent fermion masses for the first two families, since it predicts $N=U\propto D=L$, where $N,U,D,L$ are Dirac mass matrices for right-handed neutrinos, up and down quarks, and charged leptons respectively. In addition, the proportionality between the up and down quarks results in vanishing flavor (Cabibbo-Kobayashi-Maskawa) mixing \cite{Langacker:1980js}. Also the mass relations resulted from YU $m_{c}^{0}/m_{t}^{0}=m_{s}^{0}/m_{b}^{0}$ are clearly contradicted with the experimental results, where the superscript zero refers the parameters evaluated at $\mgut$. In order to correct these mass relations of the fermions, one can add new vector-like matter multiplets, which can mix with the fermions \cite{Witten:1979nr}. Another approach is to extend the content with new Higgs fields from another representations \cite{Babu:1992ia}. In this case, one can assume that the extra Higgs fields negligibly interact with the third family matter fields, and the MSSM Higgs doublets reside solely in $10-$plet of $SO(10)$ to maintain YU for the third family fermions, while the mass relations for the first two family fermions are corrected with these extra Higgs fields \cite{Joshipura:2012sr}. 

On the other hand, if we do not follow the assumptions mentioned above, the two approaches break YU. In this case, if we restrict the deviations in the Yukawa couplings of the third family up to, say  $20\%$; then, this modified unification scheme is called QYU \cite{Gomez:2002tj}. Even though the deviation is restricted to small amounts, QYU yields drastically different phenomenology at the low scale. For instance, $4-2-2$ is the only model, as to our knowledge, which yields light gluino ($m_{\tilde{g}}\lesssim 1$ TeV) to be next to lightest supersymmetric particle (NLSP), when YU is imposed at the GUT scale \cite{Gogoladze:2009ug}.  Relaxing it to $b-\tau$ YU allows stop NLSP solutions in addition to gluino \cite{Raza:2014upa}. On the other hand, QYU in $4-2-2$ allows a variety of NLSP species, while stop and gluino NLSP solutions are not compatible with QYU \cite{Dar:2011sj}. In addition, the parameter space compatible with YU yields large fine-tuning. While QYU can be realized with acceptable fine-tuning \cite{Dar:2011sj}.   

Based on the discussion above, YU provides a strict framework, in which the representations from a possible GUT gauge group is rather required to be minimal, since the MSSM Higgs fields are allowed to reside in $SO(10)$'s $10-$plet. Even though the framework can be extended in the QYU case, it is still minimalistic since only one extra representation for the Higgs fields (with those from $(15,1,3)$ \cite{Gomez:2002tj}) is allowed. However, if the framework is extended to include other possible Higgs representations, the MSSM Higgs fields become linear superpositions of those from these representations, and the  Yukawa couplings can receive different contributions depending on the vertices between the relevant matter and Higgs fields \cite{Bajc:2004xe}. In addition to the extra Higgs fields, the presence of higher dimensional operators also contribute to the Yukawa coupling such that the top quark Yukawa coupling can receive a significant correction from such operators \cite{Antusch:2013rxa}, and its deviation from YU cannot be restricted within $t-b-\tau$ QYU. In this context, the unification scheme can be identified as $b-\tau$ QYU \cite{Hebbar:2017olk}. 

The discussion followed so far does not consider the right-handed neutrinos. If LR symmetry is imposed in $4-2-2$, it requires the existence of the right-handed neutrinos, which actively participate in interactions through $SU(2)_{R}$ gauge group. Usually the effects from the right-handed neutrinos can be neglected safely due to the smallness of neutrino masses established by the experiments \cite{Wendell:2010md}, which stringently restricts the neutrino Yukawa coupling as $y_{\nu}\lesssim 10^{-7}$ \cite{Coriano:2014wxa}. On the other hand, this result does not hold when the IS mechanism is implemented, in which a large neutrino Yukawa coupling ($y_{\nu}\sim y_{t}$) can still be consistent with the smallness of neutrino masses \cite{Khalil:2010iu}.  With the presence of the right-handed neutrinos with a large Yukawa coupling, the unification scheme discussed above should be modified to include the right-handed neutrinos. In this case YU should be imposed as $y_{t}=y_{b}=y_{\tau}=y_{\nu} \equiv y$ at $\mgut$. In SUSY models, the right-handed neutrinos, in contrast to the charged leptons, interact with $H_{u}$, and the deviation in $y_{\nu}$ from YU should be proportional to those which deviate $y_{t}$ from YU. In the case of $b-\tau$ QYU, also one should impose another QYU scenario between $y_{t}$ and $y_{\nu}$ simultaneously. Following Refs. \cite{Gomez:2002tj,Dar:2011sj}, the deviations in Yukawa couplings can be formulated as 

\begin{equation}
\setstretch{1.5}
\begin{array}{l}
y_{b}:y_{\tau}=\mid 1- C_{b\tau} \mid:\mid 1+3C_{b\tau}\mid \\ 
y_{t}:y_{\nu}=\mid 1+C_{t\nu}\mid:\mid 1- 3C_{t\nu}\mid,
\label{QYUCondition}
\end{array}
\end{equation}
where $C_{b \tau}$ and $C_{t \nu}$ measure the deviation in Yukawa couplings from YU. Note that  since YU is broken first by the higher order operators as $y_{t}=y_{\nu}$ and $y_{b}=y_{\tau}$, $C_{b \tau}$ and $C_{t \nu}$ are not related to each other.

Previous studies of QYU (see for instance Refs. \cite{Gomez:2002tj,Dar:2011sj}) have revealed that the general QYU scenarios are mostly compatible in the regions with large $\tan\beta$. Such regions, depending on the mass spectrum of the supersymmetric particles, can also yield large SUSY contributions to muon $g-2$. The SM predictions exhibit about $3\sigma$ deviation from the experimental results, and this situation can be expressed as \cite{Bennett:2006fi}

\begin{equation}
 \Delta a_{\mu}\equiv a_{\mu}^{{\rm exp}}-a_{\mu}^{{\rm SM}}=(28.7\pm 8.0)\times 10^{-10} ~(1\sigma)~.
\end{equation} 

This discrepancy has been survived even after highly accurate calculations over the SM predictions were performed \cite{Davier:2010nc}; therefore, the discrepancy can be interpreted as the effect of the new physics beyond the SM. In our work, we will explore the low scale implications of $4-2-2$ including the TeV scale right-handed neutrinos, which interact and mix with the MSSM fields through the IS mechanism in the light of muon $g-2$ resolution, and highlight the solutions which are compatible with the QYU condition given in Eq. (\ref{QYUCondition}). The rest of the paper is organized as follows: We will briefly discuss the effect of the presence of the right-handed neutrinos on muon $g-2$ along with the sparticle mass spectrum in Section \ref{sec:MSSMIS}. We describe our scanning procedure and the experimental constraints employed in our data generation and analyses are summarized in Section \ref{sec:scan}. Then, we first present our results for muon $g-2$ and the sparticle spectrum in Section \ref{sec:res}. Section \ref{sec:FTg2} discusses muon $g-2$ resolution in respect of the fine-tuning, which is required to have correct electroweak symmetry breaking scale. Finally we will summarize and conclude in Section \ref{sec:conc}.

\section{Muon $g-2$ in MSSM with Inverse Seesaw}
\label{sec:MSSMIS}

\begin{figure}[h!]
\centering
\includegraphics[scale=0.3]{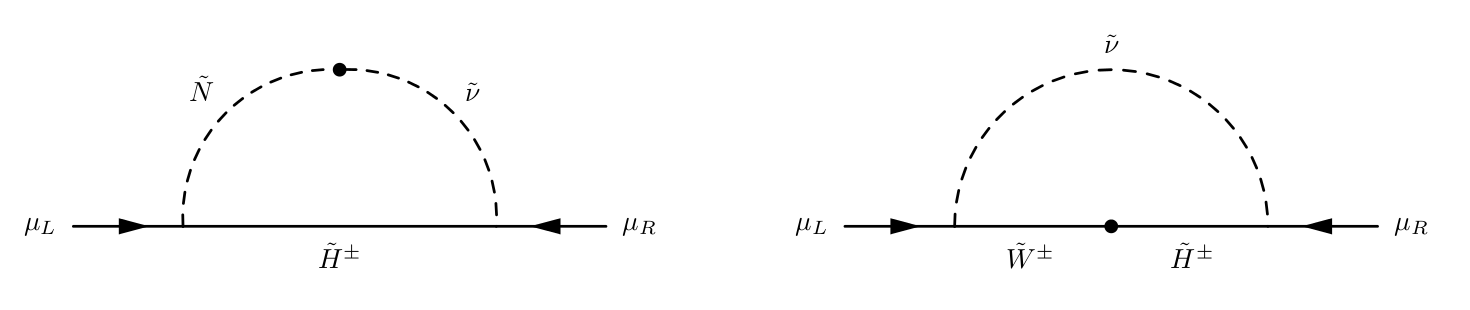}
\caption{The SUSY contributions to muon $g-2$ involving with the sneutrinos and charginos.}
\label{fig1}
\end{figure}

We will discuss the SUSY contributions to muon $g-2$ in MSSM, when the right-handed neutrinos are present and they mix through the IS mechanism. In addition to the SUSY contributions to muon $g-2$ in MSSM \cite{Moroi:1995yh}, we also present two diagrams involving with the neutrinos to illustrate the contributions arising because of the IS mechanism, in which also the charginos are running in the loop due to the charge conservation. The behavior of the SUSY contributions can be understood by calculating these diagrams with the mass insertion method, which is represented with dots in the diagrams. The approximate contributions can be obtained as follows \cite{Khalil:2015wua}:

\begin{equation}\hspace{-3.2cm}
(\Delta a_{\mu})_{C1}\approx \frac{m_{\mu}^{2}\mu^{2}\cot\beta}{m^{2}_{\tilde{N}}-m^{2}_{\tilde{\nu}}} \left[  \frac{f_{\chi}(\mu^{2}/m^{2}_{\tilde{N}})}{m^{2}_{\tilde{N}}} -\frac{f_{\chi}(\mu^{2}/m^{2}_{\tilde{\nu}})}{m^{2}_{\tilde{\nu}}}  \right]
\label{aC1}
\end{equation}

\begin{equation}\hspace{-2.8cm}
(\Delta a_{\mu})_{C2}\approx \frac{m^{2}_{\mu}M_{2}\mu\tan\beta}{m^{2}_{\tilde{N}}}  \left[ \frac{f_{\chi}(M_{2}^{2}/m^{2}_{\tilde{N}})-f_{\chi}(\mu^{2}/m^{2}_{\tilde{N}})}{M_{2}^{2}-\mu^{2}}   \right]
\label{aC2}
\end{equation}

The $\tan\beta$ dependence of muon $g-2$ contributions can be seen from these equations. Note that there are other terms, which do not depend on $\tan\beta$, but these terms are rather negligible, unless the supersymmetric particles are so light that they are excluded by the current mass bounds. The two diagrams shown in Figure \ref{fig1} exhibits different behavior in respect of $\tan\beta$. The first diagram is expected to be effective, when $\tan\beta \ll 1$. Recall that if the charged sleptons ran in the loop, the contributions would be enhanced with $\tan\beta$. The suppression in the sneutrino case is because the sneutrinos interact with $H_{u}$, while charged sleptons interact with $H_{d}$. On the other hand, the contributions represented in the second diagram exhibits an enhancement as $\tan\beta$ increases. In this processes, the $\tan\beta$ enhancement arises from the mixing between two charginos; i.e. the Wino and Higgsino. In this sense, one can expect that the second diagram illustrates the dominant processes in the SUSY contributions to muon $g-2$. 

We should note here that the second diagram is already present in the usual MSSM framework, in which the right-handed neutrinos, and hence the IS mechanism, are absent. In this context, one can conclude that the resolution to muon $g-2$ discrepancy is not improved too much when the right-handed neutrinos are present and they mix through the IS mechanism. However, a recent study shows that muon $g-2$ discrepancy can be significantly resolved in MSSM extended with a $U(1)_{B-L}$ symmetry (BLSSM) \cite{Khalil:2015wua} when the IS mechanism is also implemented. The resolution happens even the universal boundary conditions are imposed at $\mgut$, which is not possible in the MSSM framework without the IS mechanism. Thus, the effects of the right-handed sneutrinos could be indirect, while their direct contributions to muon $g-2$ is significantly suppressed by $\tan\beta$. Such indirect effects can be understood by considering the following RGEs for the relevant parameters, which run the parameters from $\mgut$ to the low scale

\begin{equation}
\setstretch{2.5}
\begin{array}{ll}
\dfrac{dm_{\tilde{L}}^{2}}{dt} & =\left( \dfrac{dm_{\tilde{L}}^{2}}{dt} \right)_{MSSM}-2m_{H_{u}}^{2}Y_{\nu}^{\dagger}Y_{\nu}-2T_{\nu}^{\dagger}T_{\nu}-2m_{\tilde{l}}^{2}Y_{\nu}^{\dagger}Y_{\nu} -2Y_{\nu}^{\dagger}m_{\tilde{\nu}}^{2}Y_{\nu} \\
\dfrac{dm_{\tilde{e}}^{2}}{dt} & =\left( \dfrac{dm_{\tilde{e}}^{2}}{dt} \right)_{MSSM} \\
\dfrac{dm_{\tilde{N}}^{2}}{dt} & =-2 \left( 2m_{H_{u}}^{2} Y_{\nu}Y_{\nu}^{\dagger}+2 T_{\nu}T_{\nu}^{\dagger}+2 Y_{\nu}m_{\tilde{l}}^{2}Y_{\nu}^{\dagger} +2m_{\tilde{\nu}}^{2}Y_{\nu}Y_{\nu}^{\dagger} \right) \\
\dfrac{dm_{H_{u}}^{2}}{dt}& =\left( \dfrac{dm_{H_{u}}^{2}}{dt} \right)_{MSSM}-2m_{H_{u}}^{2}Tr\left(Y_{\nu}Y_{\nu}^{\dagger}\right)-2Tr\left(T_{\nu}^{\ast}T_{\nu}^{T}\right) \\
& \hspace{3.4cm}-2Tr\left(m_{\tilde{l}}^{2}Y_{\nu}^{\dagger}Y_{\nu}\right)-2Tr\left( m_{\tilde{\nu}}^{2}Y_{\nu}Y_{\nu}^{\dagger} \right) \\
\dfrac{d\mu}{dt} & =\left( \dfrac{d\mu}{dt} \right)_{MSSM}-\mu Tr \left( Y_{\nu}Y_{\nu}^{\dagger} \right)
\end{array}
\label{RGEs}
\end{equation}
where we have used the usual notation for the MSSM fields. In addition, $\tilde{N}$ denotes the right-handed sneutrino field. The first terms in the equations with the subscript MSSM represent the RGEs for these parameters within the MSSM framework without the right-handed neutrino. The other terms are relevant to the presence of the right-handed neutrinos. As is seen from the RGEs given above, the neutrino Yukawa couplings, $Y_{\nu}$, and its trilinear interaction term $T_{\nu}$ are effective in lowering the SSB masses of the charged sleptons, and as a result they could be much lighter at the low scale than those in the usual MSSM without the right-handed neutrinos. If the texture of the neutrino Yukawa couplings are similar to the up-type quarks Yukawa couplings ($Y_{\nu}\sim Y_{u}$) \cite{Khalil:2010iu}, then neutrino Yukawa couplings and trilinear interaction term can result in tachyonic states ($m_{L,e,N}^{2} < 0$) especially for staus. In this context, the smuon masses can be found slightly heavier in order to avoid tachyonic stau mass eigenstates \cite{Gogoladze:2014vea}. However, the SUSY contributions to muon $g-2$ from smuon-neutralino loop can be suppressed, if $m_{\tilde{\mu}} \gtrsim 800$ \cite{Babu:2014lwa}.

Similar discussion holds for the right-handed sneutrinos. Its SSB mass parameter is determined with the common mass scale for the scalars, $m_{0}$; and hence, $m_{0}$ cannot be lower than certain scales not to have tachyonic sneutrinos at the low scale. The RGE for $m_{H_{u}}$ reveals an interesting feature for the IS mechanism that $Y_{\nu}$ and $T_{\nu}$ lower its value from $\mgut$ to the low scale as $Y_{t}$ gives the same impact as MSSM. The electroweak symmetry breaking requires $m_{H_{u}}^{2} < 0$, and MSSM can have only stops to have negative $m_{H_{u}}^{2}$ at the low scale. This fact leads to heavy stops and/or large mixing between left and right-handed stops in the MSSM. On the other hand, when the IS mechanism is implemented in the MSSM framework, the sneutrinos, together with the stops, yield $m_{H_{u}}^{2} < 0$, which loose the pressure on the stop sector. In this case, even if the mixing between the left and right-handed stops are small, it is still possible to have stops at around TeV scale in the mass spectra. 

Before concluding this section, the last RGE for the $\mu-$term in Eq.(\ref{RGEs}) is also interesting in the naturalness point of view. As shown in previous studies \cite{Baer:2012mv}, the required fine-tuning at the electroweak scale is mostly determined by $\mu$. Its RGE in the case with the IS mechanism shows that the $\mu-$term is lowered further than that in the usual MSSM by the neutrino Yukawa couplings and trilinear scalar interaction terms. Hence, one can expect that the MSSM with IS can yield significantly low fine-tuned solutions at the low scale. 

Note that even though the RGEs are more or less the same in the case with Type I seesaw, the terms with neutrino Yukawa couplings and trilinear scalar interaction term are quite negligible, since $Y_{\nu}\lesssim 10^{-7}$. In this sense, despite the presence of the right-handed neutrinos, the low scale implications of SUSY Type-I Seesaw are almost the same as those in the usual MSSM models.

\section{Scanning Procedure and Experimental Constraints}
 \label{sec:scan}

In scanning the fundamental parameter space, we have employed SPheno 3.3.3 package \cite{Porod:2003um} obtained with SARAH 4.6.0 \cite{Staub:2008uz}. This package evolves the weak scale values of gauge and Yukawa couplings to $\mgut$ via the MSSM RGEs, which are modified to include the IS mechanism. $\mgut$ is dynamically determined with the gauge coupling unification condition. Note that we do not strictly  enforce the unification condition at $\mgut$, since a few percent deviation from the unification can be assigned to unknown GUT-scale threshold corrections \cite{Hisano:1992jj}, which modify the unification condition as $g_{1}=g_{2}\approx g_{3}$. With the boundary conditions given at $M_{{\rm GUT}}$, all the SSB parameters along with the gauge and Yukawa couplings are evolved back to the weak scale.

We have performed random scans over the following parameter space 

\begin{equation}
\setstretch{1.5}
\begin{array}{ccc}
0 \leq &  ~ m_{0},m_{H_{d}},m_{H_{u}} ~  &  \leq ~ 5~{\rm TeV}, \\
-5 \leq &  ~ M_{2}~&\leq ~ 0~{\rm TeV}, \\
0 \leq &  ~ 	M_{3}~&\leq ~ 5~{\rm TeV}, \\
-3  \leq & A_{0}/m_{0} &\leq 3  \\
 35  \leq & \tan\beta & \leq 60
 \end{array}
\label{paramSpace}
\end{equation}
\begin{equation*}
\mu < 0~,\hspace{0.3cm} m_{t}=173.3~{\rm GeV}
\end{equation*}
where $m_{0}$ symbolizes the SSB mass term for the matter scalars, while $m_{H_{d}}$ and $m_{H_{u}}$ denote the SSB mass terms for the MSSM Higgs doublets. $M_2$ and $M_3$  stand for the gauginos associated with the $SU(2)_{L}$ and $SU(3)_{C}$ respectively. The SSB mass term, $M_{1}$ for the $U(1)_{Y}$ gaugino is determined in terms of $M_{2}$ and $M_{3}$ as given in Eq.(\ref{422gauginos}). $A_0$ is the SSB trilinear coupling, and $\tan\beta$ is the ratio of vacuum expectation values (VEVs) of the MSSM Higgs doublets. The value of $\mu$-term in MSSM is determined by the radiative electroweak symmetry breaking (REWSB) condition but not its sign; thus, its sign is one of the free parameter in MSSM and it is set negative in our scans. In addition, we have employed the current central value for the top quark mass as $m_t=173.3$ GeV \cite{Group:2009ad}. Note that the sparticle spectrum is not too sensitive in one or two sigma variation in the top quark mass \cite{Gogoladze:2011db}, but it can shift the Higgs boson mass by 1-2 GeV \cite{Gogoladze:2011aa}.  In addition to these free parameters, the experiments do not provide any value for the neutrinos Yukawa coupling at the low scale, in contrast to those associated with the charged leptons. Hence, they need to be provided as an input at the low scale. In our scans, we vary Yukawa coupling $Y_{\nu}$ within perturbative level.

In adjusting the ranges of the free parameters, we restrict the scalar and gaugino SSB mass terms not to exceed 5 TeV in order to remain in the regions which yield acceptable fine-tuning at the low scale. The range of the trilinear scalar coupling is set to avoid charge and/or color breaking minima, which requires $|A_{0}|\lesssim 3m_{0}$. Among these parameters, we bound $\tan\beta$ at 35 from below. Even though the general MSSM framework can be consistent with the current experimental results including the Higgs boson mass, Yukawa unification requires rather large $\tan\beta$ to satisfy the correct masses for quarks and charged leptons \cite{bigger-422}. Even in the case of QYU, the unification scheme needs $\tan\beta \gtrsim 40$ \cite{Dar:2011sj}.

The REWSB condition puts crucial theoretical constraint \cite{Ibanez:Ross} on the parameter space given in Eq.(\ref{paramSpace}). According to this constraint, the SSB mass-squared terms for the Higgs doublets are required to be negative at the low scale, though they are positive-defined at $\mgut$. In this context, the relevant parameters in the RGE evolutions of these mass parameters should be tuned in a way that, $m_{H_{u}}^{2}$ and/or $m_{H_{d}}^{2}$ have to be turn negative. Another constraint is dark matter observations and it restricts the parameter space which requires the lightest supersymmetric particle (LSP) stable and no electric and color charge, which excludes the regions leading to stau or stop LSP solutions \cite{Nakamura:2010zzi}. On the other hand, even if a solution does not satisfy the dark matter observations, it can still survive in conjunction with other form(s) of the dark matter formation \cite{Baer:2012by}. Based on this discussion, we accept only the solutions which yield neutralino LSP at the low scale, but we do not apply any constraint from the dark matter experiments.

In scanning the parameter space, we use our interface which employs Metropolis-Hasting algorithm described in \cite{Belanger:2009ti}. All collected data points satisfy the requirement of REWSB and neutralino LSP. After collecting data, we subsequently impose the mass bounds on all 	the sparticles \cite{Agashe:2014kda} and the constraints from rare decay processes $B_s \rightarrow \mu^+ \mu^-$ \cite{Aaij:2012nna} and $b \rightarrow s \gamma$ \cite{Amhis:2012bh}. In addition those bounds we have imposed Higgs boson \cite{Aad:2012tfa} and gluino masses \cite{TheATLAScollaboration:2015aaa}. The experimental constraints mentioned above can be summarized below:
\begin{equation}
\setstretch{1.5}
\begin{array}{rr}
m_{\tilde{\chi}^{\pm}_{1}} & \geq  103.5~{\rm GeV}, \\
 123 \leq m_h & \leq ~ 127~~{\rm GeV}, \\
m_{\tilde \tau} & \geq ~ 105~{\rm GeV}, \\
m_{\tilde g} & \geq  1800~{\rm GeV}, \\
 \end{array}
\label{experimentalcons}
\end{equation}
\begin{equation*}
\begin{split}
0.8\times 10^{-9} \leq BR(B_s \rightarrow \mu^+ \mu^-)~~ & \leq \quad 6.2 \times 10^{-9} (2\sigma), \\
2.99\times 10^{-4} \leq BR(b \rightarrow s \gamma) ~~ & \leq \quad 3.87 \times 10^{-4} (2\sigma) \\
\end{split}
\end{equation*}

Finally we identify the regions compatible with QYU by restricting the deviations in the Yukawa couplings within to $20\%$ by applying $C_{t\nu} \leq 0.2$ and $C_{b\tau}\leq 0.2$, which refers to the QYU condition.

\section{Fundamental Parameter Space of Muon $g-2$ and Sparticle Spectrum}
\label{sec:res}

\begin{figure}[ht!]
\centering
\subfigure{\includegraphics[scale=1]{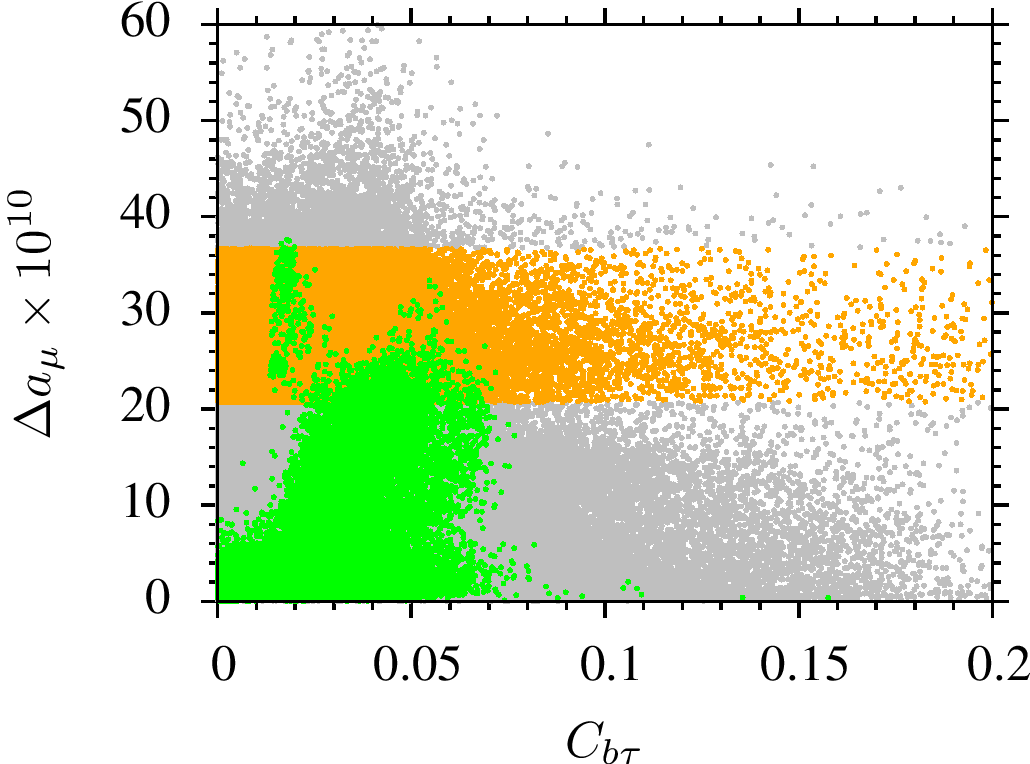}}
\subfigure{\includegraphics[scale=1]{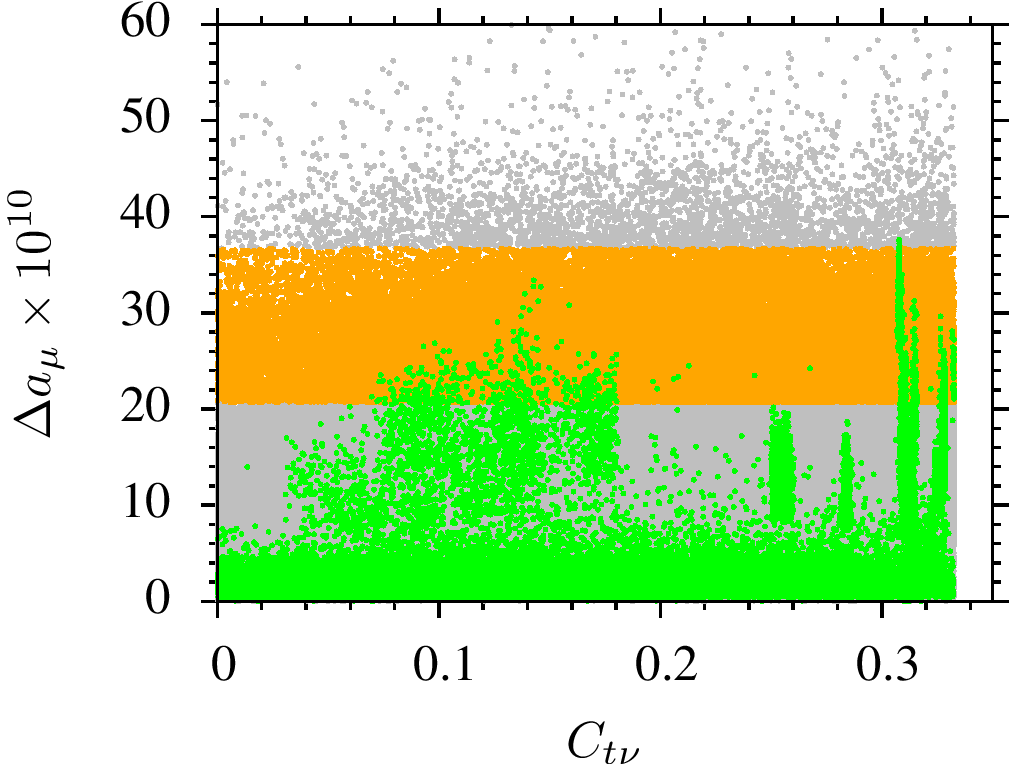}}
\caption{Plots in the $\Delta a_{\mu}-C_{b\tau}$ and $\Delta a_{\mu}-C_{t\nu}$ planes. All points are consistent with REWSB and neutralino LSP. Green points satisfy the experimental constraints mentioned in Section \ref{sec:scan}. Yellow band is an independent subset of gray points, and they indicate the values of $\Delta a_{\mu}$ which would bring the theory and the experiment within $1\sigma$.}
\label{fig2}
\end{figure}

\begin{figure}[ht!]
\centering
\subfigure{\includegraphics[scale=1]{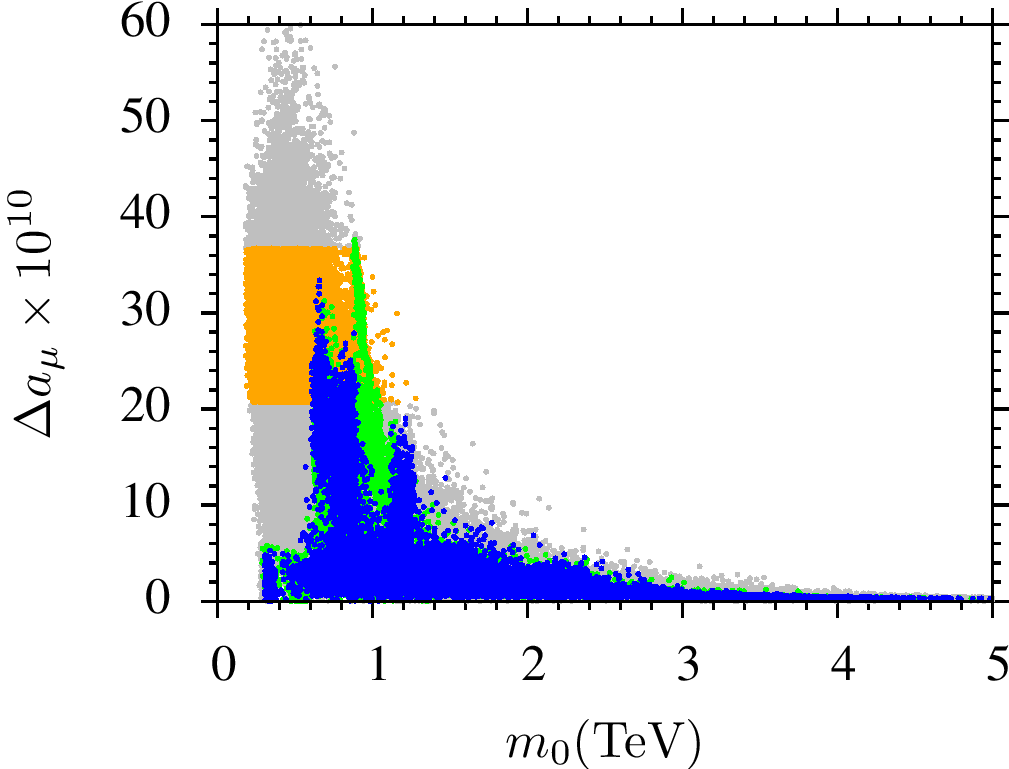}}
\subfigure{\includegraphics[scale=1]{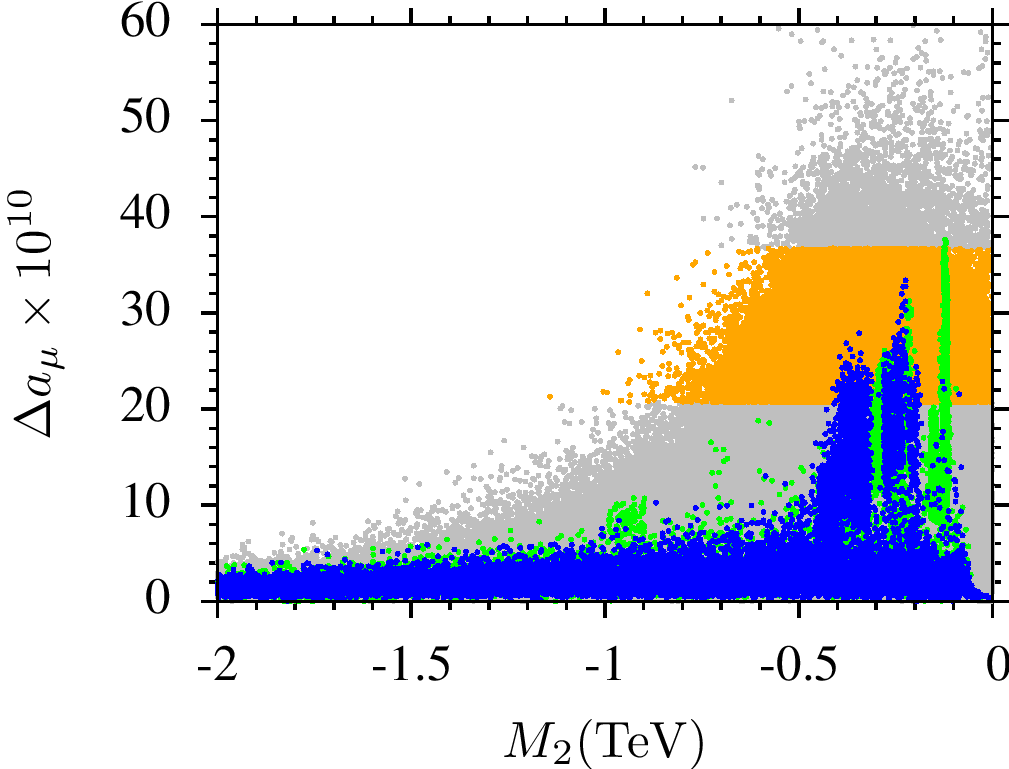}}
\subfigure{\includegraphics[scale=1]{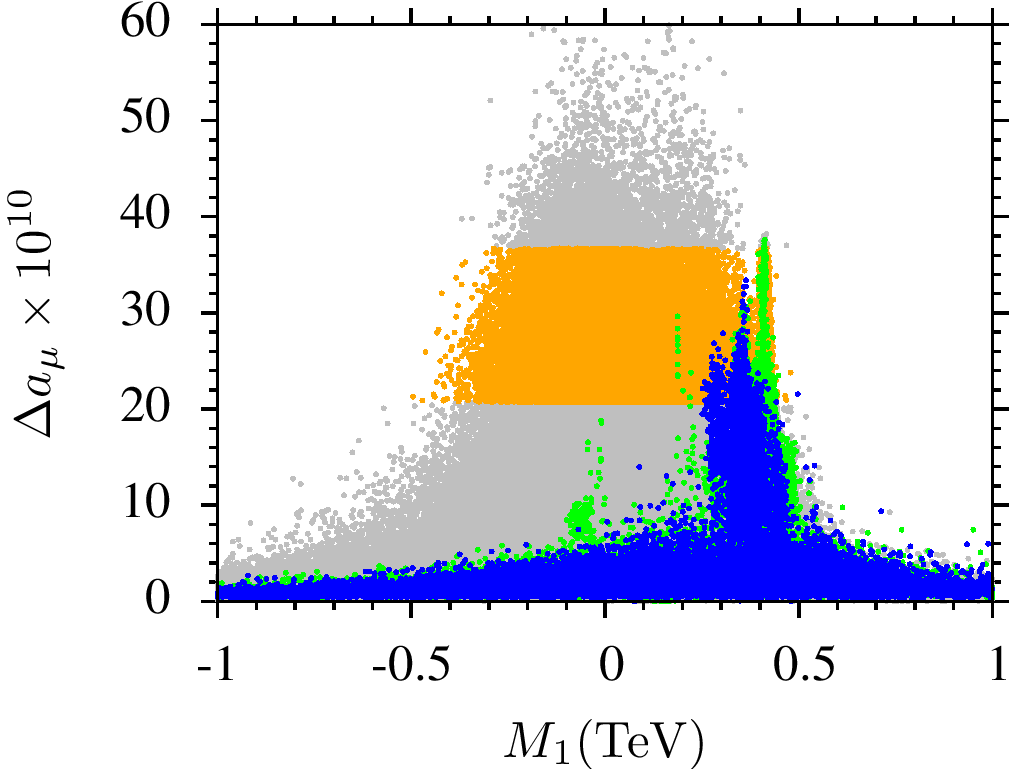}}
\subfigure{\includegraphics[scale=1]{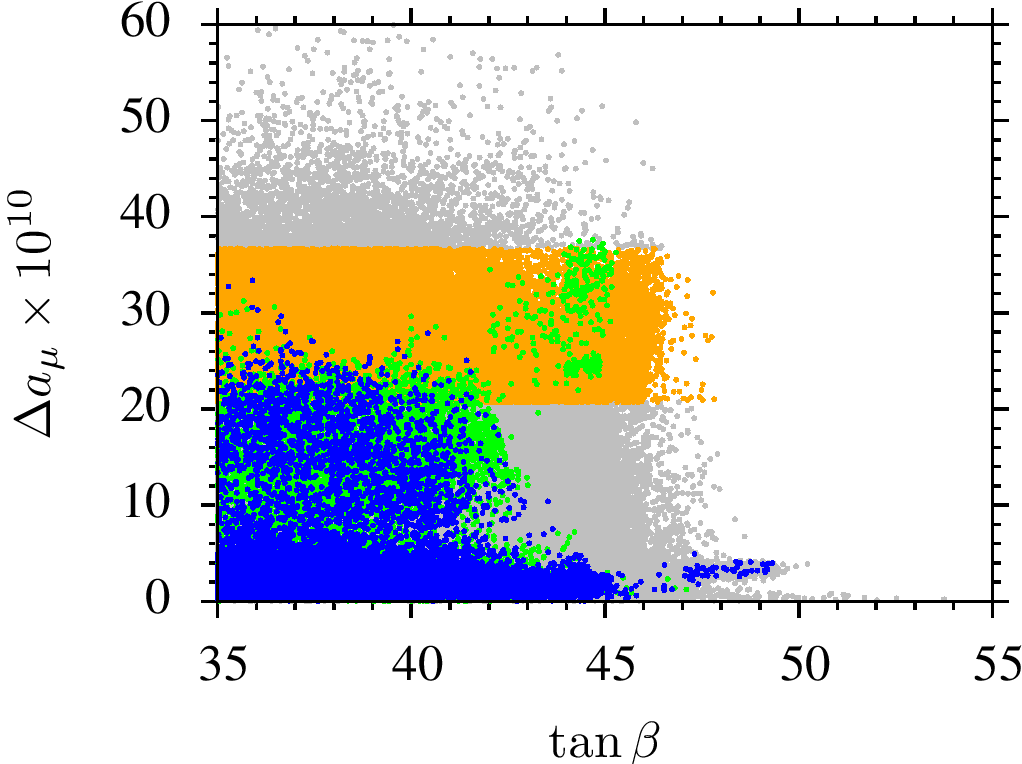}}
\caption{Plots in the $\Delta a_{\mu}-m_{0}$, $\Delta a_{\mu}-M_{2}$, $\Delta a_{\mu}-M_{1}$ and $\Delta a_{\mu}-\tan\beta$ planes. The color coding is the same as Figure \ref{fig2}. In addition, blue points form a subset of green and they represent the solutions compatible with the QYU condition.}
\label{fig3}
\end{figure}

\begin{figure}[ht!]
\centering
\subfigure{\includegraphics[scale=1]{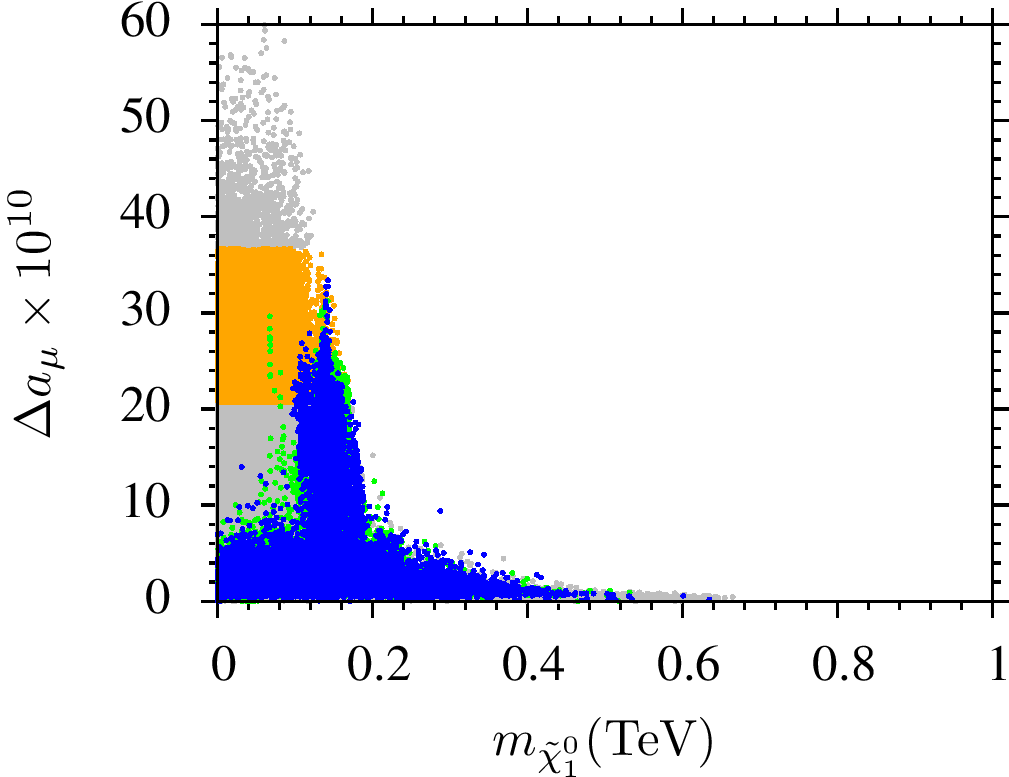}}
\subfigure{\includegraphics[scale=1]{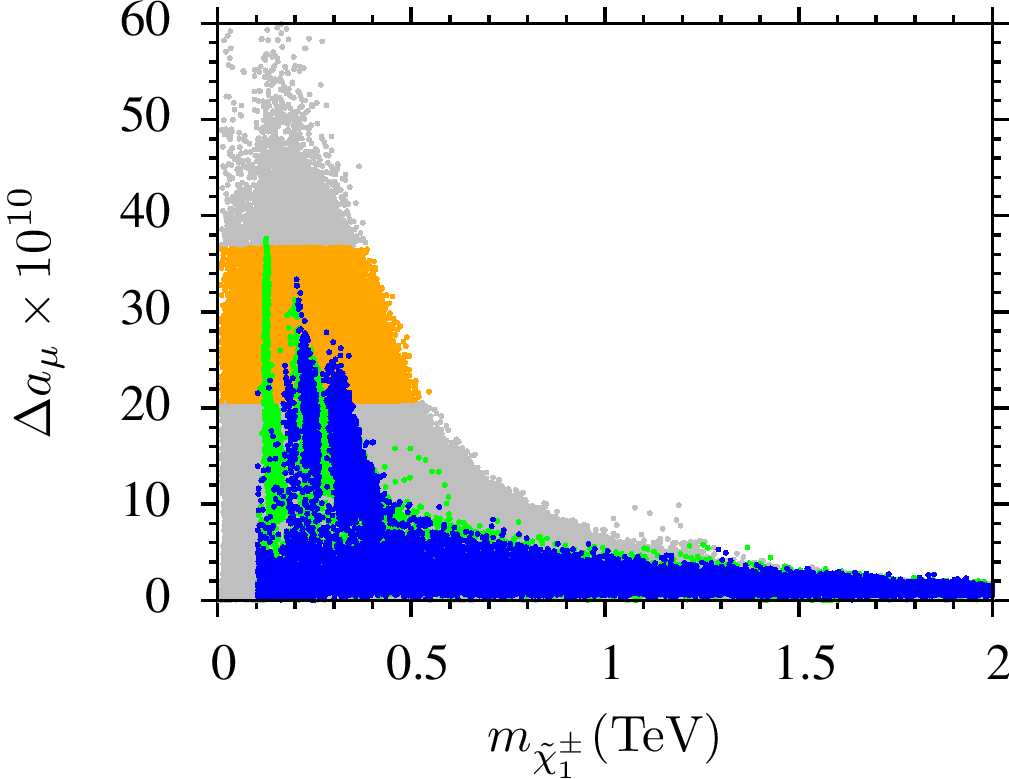}}
\subfigure{\includegraphics[scale=1]{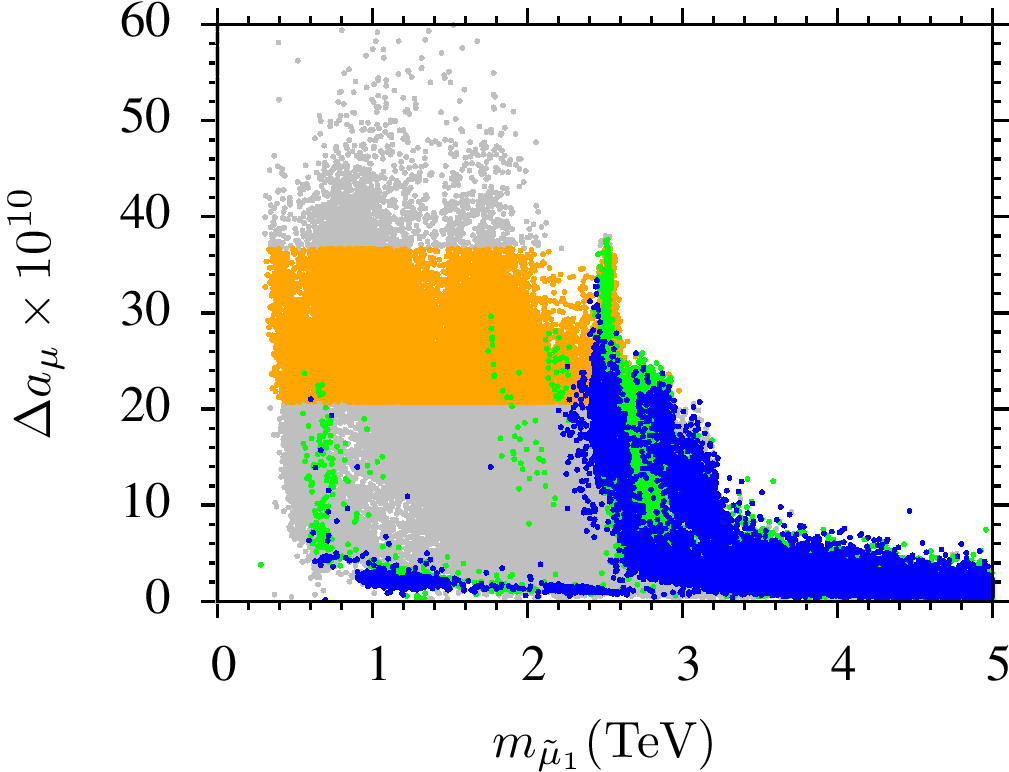}}
\subfigure{\includegraphics[scale=1]{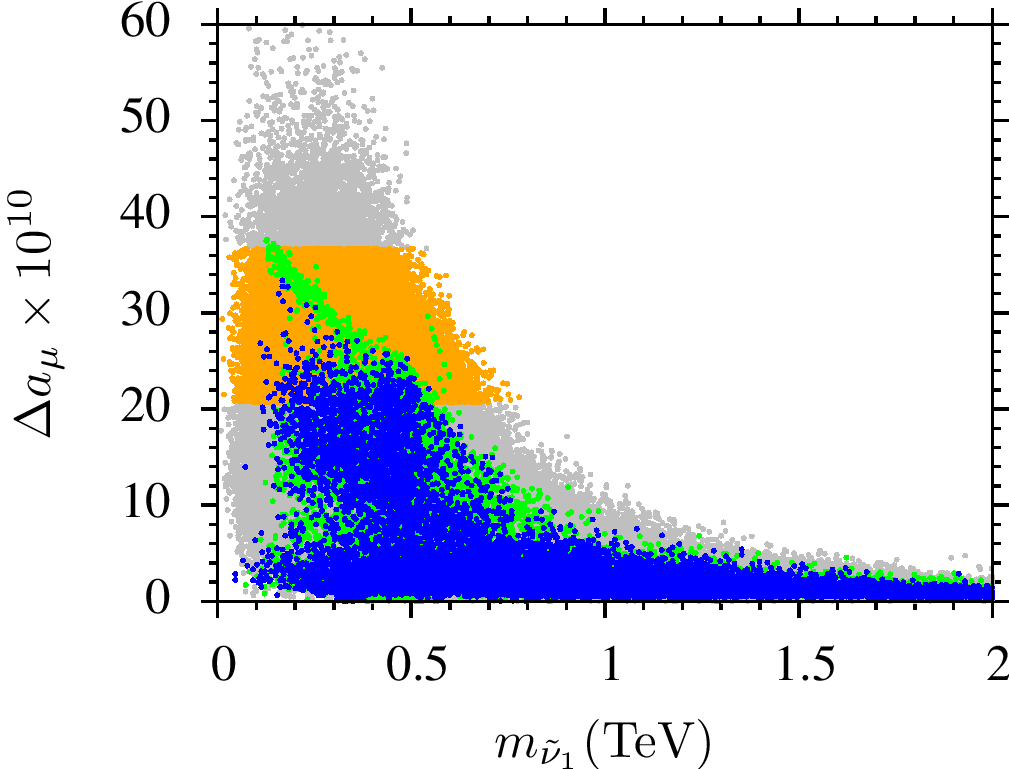}}
\caption{Plots in the $\Delta a_{\mu}-m_{\tilde{\chi}_{1}^{0}}$, $\Delta a_{\mu}-m_{\tilde{\chi}_{1}^{\pm}}$, $\Delta a_{\mu}-m_{\tilde{\mu}_{1}}$ and $\Delta a_{\mu}-m_{\tilde{\nu}_{1}}$ planes. The color coding is the same as Figure \ref{fig3}.}
\label{fig4}
\end{figure}

In this section, we discuss muon $g-2$ results and highlight the solutions compatible with the QYU condition. We start first with Figure \ref{fig2} displaying the deviations of the Yukawa couplings from the unification with plots in the $\Delta a_{\mu}-C_{b\tau}$ and $\Delta a_{\mu}-C_{t\nu}$ planes. All points are consistent with REWSB and neutralino LSP. Green points satisfy the experimental constraints mention in Section \ref{sec:scan}. Yellow band is an independent subset of gray points, and they indicate the values of $\Delta a_{\mu}$ which would bring the theory and the experiment within $1\sigma$. As seen from the $\Delta a_{\mu}-C_{b\tau}$, $C_{b\tau}$ measuring the deviation in $y_{b}$ and $y_{\tau}$ can barely reach to $20\%$, and the experimental constraints restrict it to $C_{b\tau} \lesssim 0.1$ (green). In addition, the region compatible with resolution to muon $g-2$ discrepancy (yellow band) bounds it further to $C_{b\tau} \lesssim 0.07$. On the other hand, $C_{t\nu}$ can be as large as about $0.3$ as shown in the $\Delta a_{\mu}-C_{t\nu}$. This is not surprising, since $y_{t}$, and consequently $y_{\nu}$, can receive large corrections from the extra Higgs fields and also the higher dimensional operators. However, it is still possible to restrict it within to $20\%$. Besides, imposing the QYU condition will exclude the solutions with $C_{t\nu} > 0.2$.

Figure \ref{fig3} represents the correlations between muon $g-2$ and the relevant fundamental parameters with plots in the $\Delta a_{\mu}-m_{0}$, $\Delta a_{\mu}-M_{2}$, $\Delta a_{\mu}-M_{1}$ and $\Delta a_{\mu}-\tan\beta$ planes. The color coding is the same as Figure \ref{fig2}. In addition, blue points form a subset of green and they represent the solutions compatible with the QYU condition. The $\Delta a_{\mu}-m_{0}$ shows that $m_{0}$ cannot be greater than about 1.2 TeV in order for the resolution to muon $g-2$ discrepancy. Since $m_{0}$ controls the scalar masses, it is understandable with the need of light scalars, which run in the loops contributing to muon $g-2$, at the low scale. However, the regions with $m_{0} \lesssim 600$ GeV cannot provide a resolution to muon $g-2$ consistently accommodated with the current experimental constraints. This result arises from the effects of the right-handed neutrino sector discussed along with the RGEs in the previous section. The gray region coinciding with the yellow band yield inconsistently light charged sleptons ($m_{\tilde{l}} < 100 $ GeV) especially for those from the third family. In addition, the Higgs boson mass is problematic in these regions, since most of the solutions predict $m_{h} < 125$ GeV. Similarly, muon $g-2$ condition requires light weakinos (Bino and Wino), and $|M_{2}| \lesssim 500$ GeV as seen from the $\Delta a_{\mu}-M_{2}$ plane. This parameter controls the wino mass at the low scale as $m_{\tilde{W}}\approx |M_{2}|$; hence, muon $g-2$ condition necessitates light charginos at the low scale. Similarly, $M_{1}$, which controls the Bino mass as $m_{\tilde{B}} \approx 0.5 |M_{1}|$ \cite{Gogoladze:2009bd}, needs to be light ($\lesssim 500$ GeV). Since we set in $\mu < 0$, one can expect to have large SUSY contributions to muon $g-2$ when $M_{1}$ is negative in contrast to the results shown in the $\Delta a_{\mu}-M_{1}$ plane, where the SUSY contributions seem suppressed when $M_{1} < 0$. 

We consider the low scale mass spectrum for the supersymmetric particles in Figure \ref{fig4} with plots in the $\Delta a_{\mu}-m_{\tilde{\chi}_{1}^{0}}$, $\Delta a_{\mu}-m_{\tilde{\chi}_{1}^{\pm}}$, $\Delta a_{\mu}-m_{\tilde{\mu}_{1}}$ and $\Delta a_{\mu}-m_{\tilde{\nu}_{1}}$ planes. The color coding is the same as Figure \ref{fig3}. The neutralino LSP mass cannot exceed about $200$ GeV in order for maintaining the resolution to muon $g-2$, while the lightest chargino can be as heavy as about $400$ GeV as seen from the $\Delta a_{\mu}-m_{\tilde{\chi}_{1}^{\pm}}$. In contrast to the usual MSSM implications, the IS mechanism yields rather heavy smuons ($m_{\tilde{\mu}} \gtrsim 800$ GeV), which can significantly suppress the SUSY contributions from the smuon-neutralino loop processes. As discussed above, the smuon masses are mostly bounded from below by the stau mass, which can turn to be tachyonic when the other charged sleptons are light due to its large trilinear SSB term ($T_{\tau}\equiv A_{\tau}y_{\tau}$). The solutions in such regions are required to yield $m_{\tilde{\tau}} \geq 105$ GeV to be consistent the mass bounds on sparticles. While the stau mass bound has a strong impact on the smuon masses, the sneutrinos, on the other hand, can be as light as about $100$ GeV, which yield significant SUSY contributions to muon $g-2$ along with light charginos. Consequently, the main contributions to muon $g-2$ are provided from the sneutrino-chargino loop processes, while those from smuon and neutralino are highly suppressed due to the heavy smuon masses. This also explains why there is no significant muon $g-2$ contributions when $M_{1}$ is negative. In this region, $M_{2}$ needs to be larger than $M_{3}$, which yields relatively heavy charginos at the low scale, so the SUSY contributions from chargino and sneutrino are also suppressed. In addition, the reason why the neutralino mass is bounded from above as $m_{\tilde{\chi}_{1}^{0}} \lesssim 200$ GeV is only the condition which requires neutralino to be LSP for all the solutions.

\section{Fine-Tuning and Muon $g-2$ in MSSM with IS}
\label{sec:FTg2}

\begin{figure}[ht!]
\centering
\subfigure{\includegraphics[scale=1]{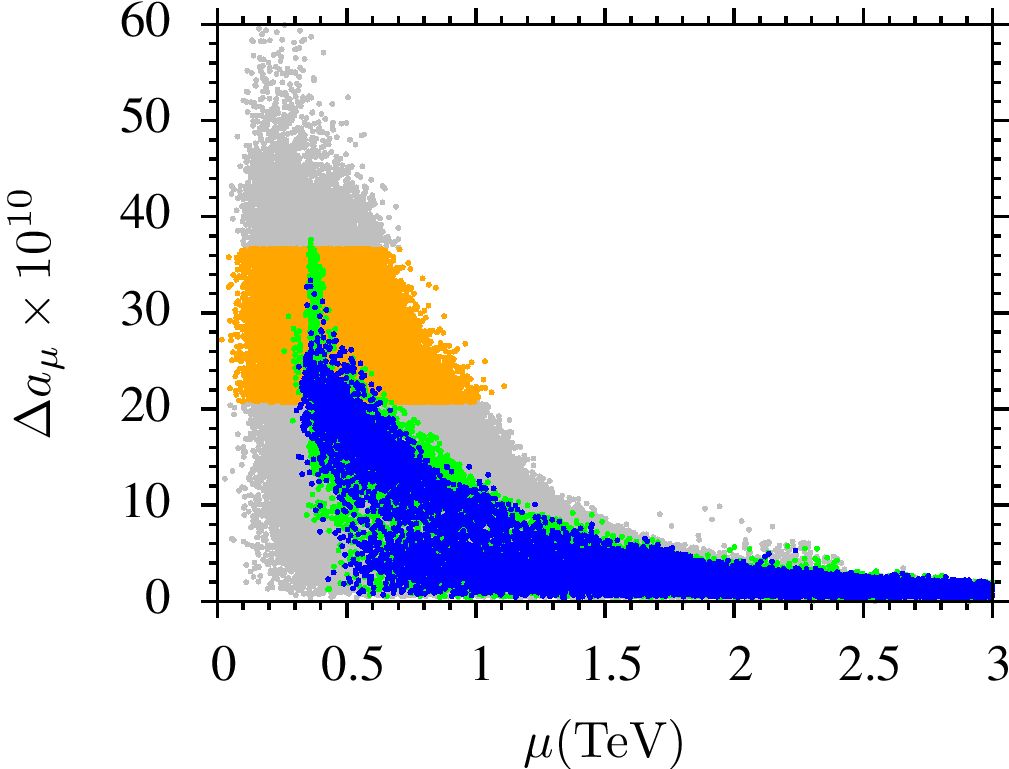}}
\subfigure{\includegraphics[scale=1]{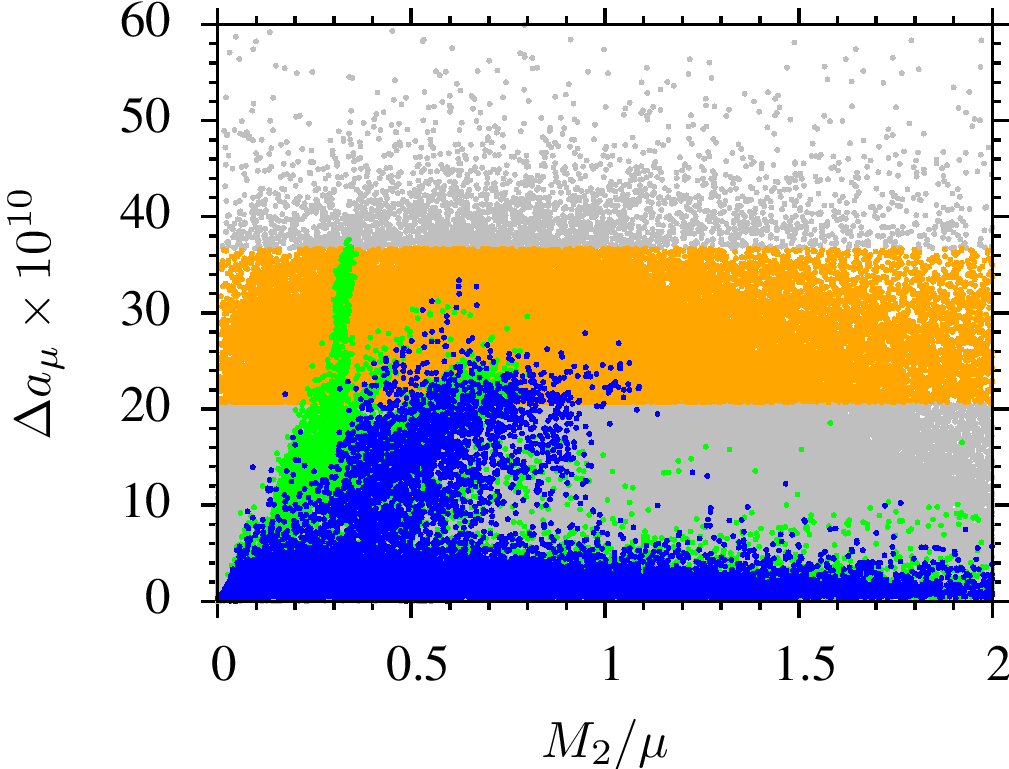}}
\subfigure{\includegraphics[scale=1]{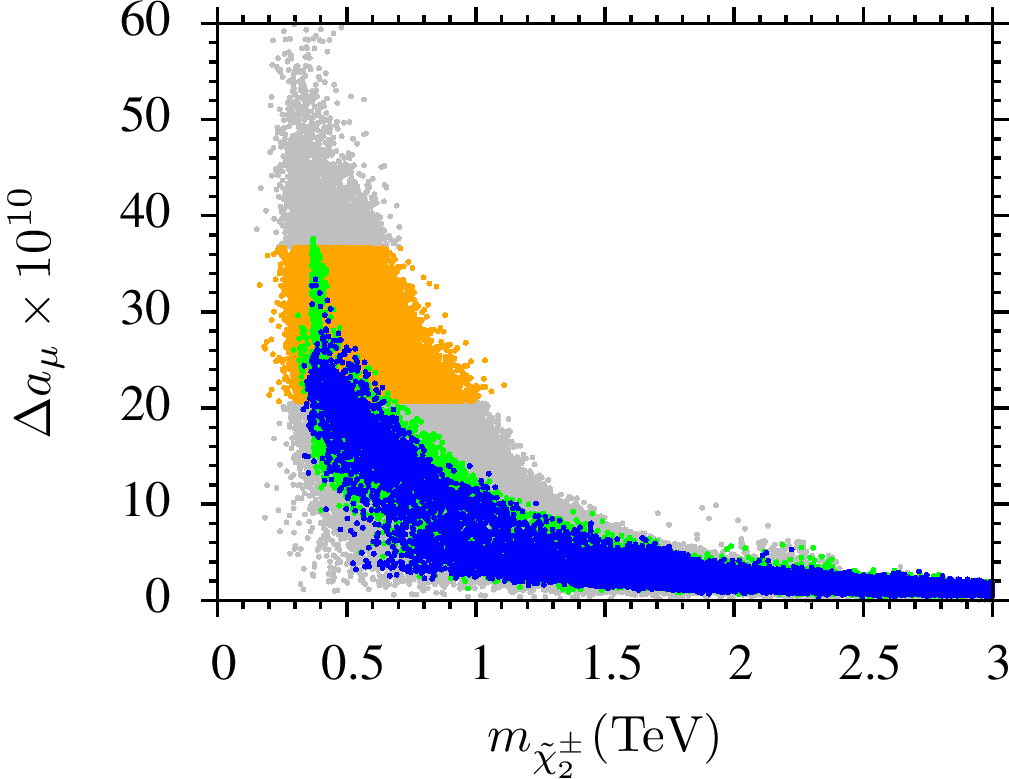}}
\subfigure{\includegraphics[scale=1]{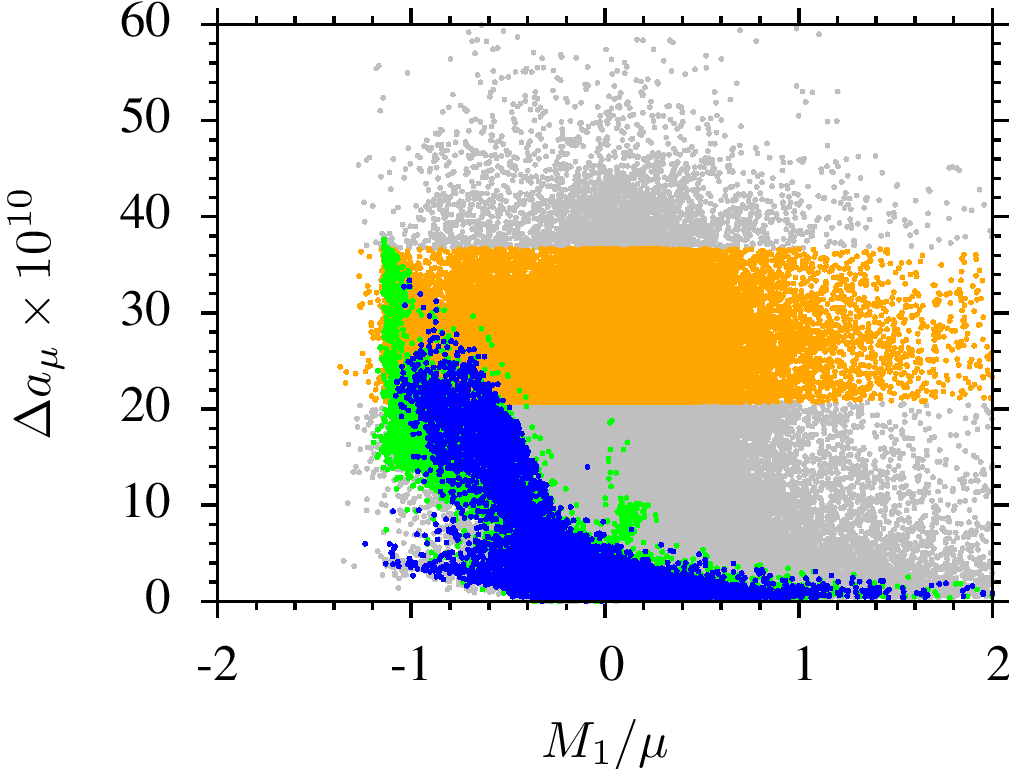}}
\caption{Plots in the $\Delta a_{\mu}-\mu$, $\Delta a_{\mu}-M_{2}/\mu$, $\Delta a_{\mu}-m_{\tilde{\chi}_{2}^{\pm}}$, and $\Delta a_{\mu}-M_{1}/\mu$ planes. The color coding is the same as Figure \ref{fig3}.}
\label{fig5}
\end{figure}


\begin{figure}[ht!]
\centering
\subfigure{\includegraphics[scale=1]{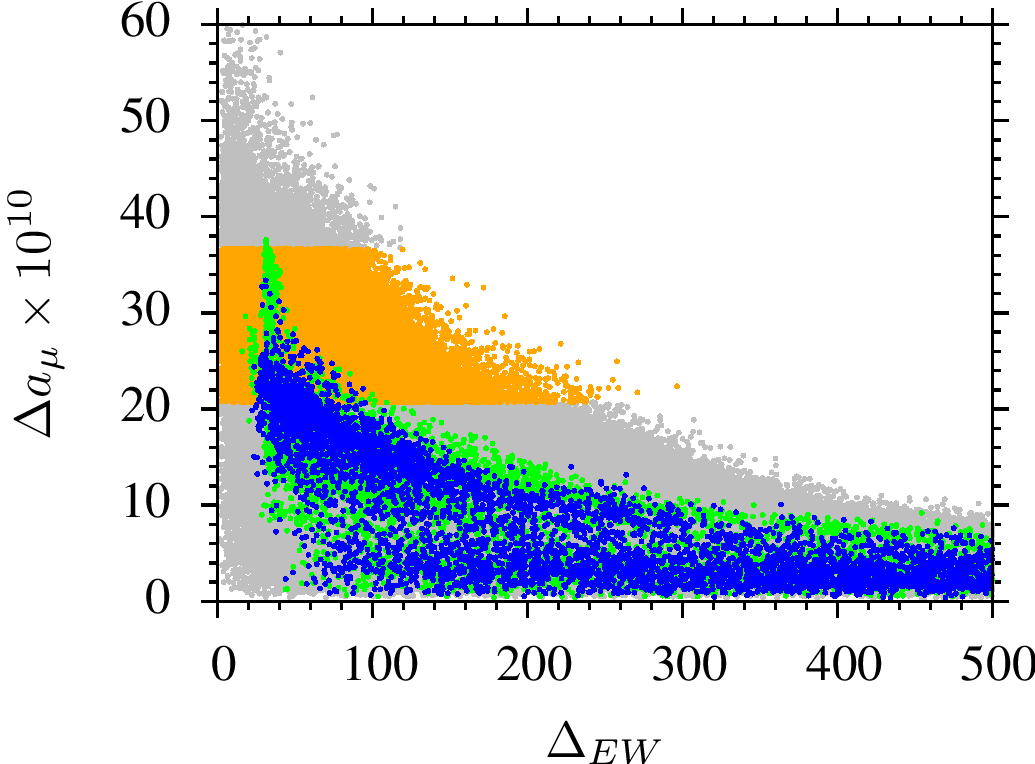}}
\subfigure{\includegraphics[scale=1]{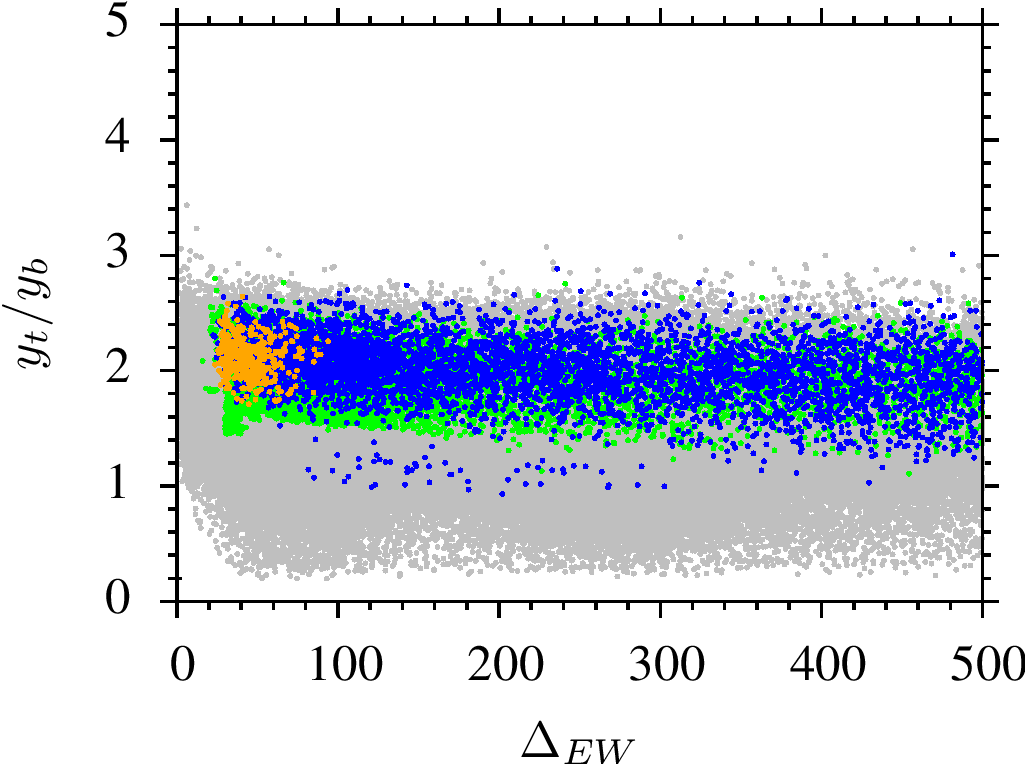}}
\subfigure{\includegraphics[scale=1]{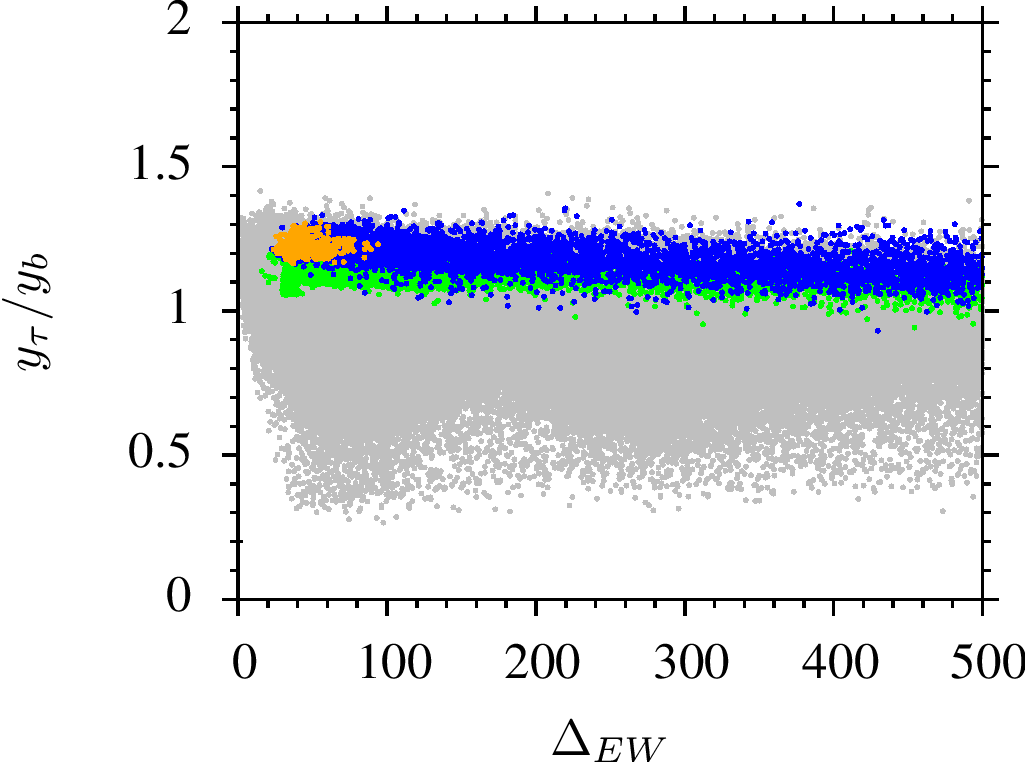}}
\subfigure{\includegraphics[scale=1]{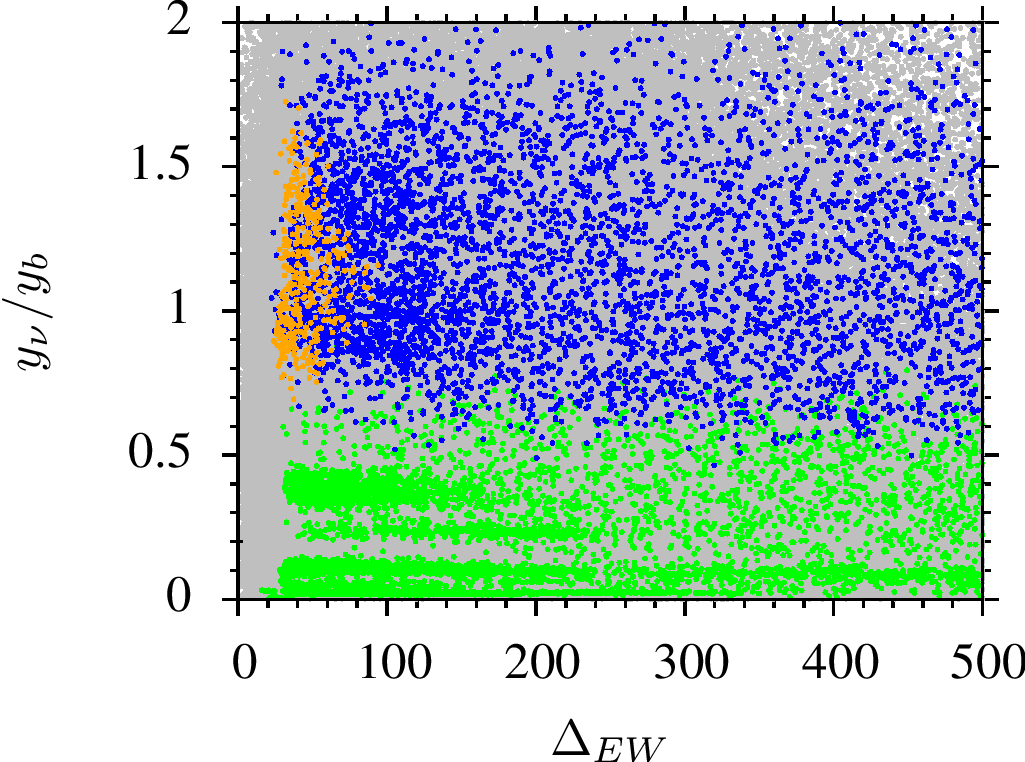}}
\caption{Plots in the $\Delta a_{\mu}-\Delta_{EW}$, $y_{t}/y_{b}-\Delta_{EW}$, $y_{\tau}/y_{b}-\Delta_{EW}$ and $y_{\nu}/y_{b}-\Delta_{EW}$ planes. The color coding is the same as Figure \ref{fig3}.}
\label{fig6}
\end{figure}


As we discussed in the previous section, the dominant contribution to muon $g-2$ comes from the sneutrino-chargino loop processes. In these processes, the chargino can be either Wino or Higgsino, each of which corresponds to different nature of the SUSY contributions. If the chargino is Wino-like, then the contributions are generated through $SU(2)$ interactions, while the Yukawa interactions take part when the chargino is Higgsino like. Depending on the ratios of their masses, the chargino could also be a mixture of these two particles. Figure \ref{fig5} represents the result for the Higgsino mass and its mass ratio to the chargino mass with plots in the $\Delta a_{\mu}-\mu$, $\Delta a_{\mu}-M_{2}/\mu$, $\Delta a_{\mu}-m_{\tilde{\chi}_{2}^{\pm}}$, and $\Delta a_{\mu}-M_{1}/\mu$ planes. The color coding is the same as Figure \ref{fig3}. According to the $\Delta a_{\mu}-\mu$ plane, the Higgsinos, whose masses are equal to $\mu$, can be as light as about 500 GeV, while muon $g-2$ condition bounds its mass from above at about 700 GeV. In this sense, the model predicts relatively light Higgsinos at the low scale compatible with the QYU condition. The $\Delta a_{\mu}-M_{2}/\mu$ plane compares the Wino and Higgsino masses to each other by considering their mass ratio. The results in this plane show that, despite the light Higgsinos, the Wino is mostly lighter than the Higgsino over all the parameter space, when the solutions yield muon $g-2$ results that would bring the experiment and theory within to $1\sigma$, since $M_{2}/\mu \lesssim 1$. On the other hand, muon $g-2$ resolution also bounds this mass ratio from below at about 0.5, which leads to a comparable mixing between the Wino and the Higgsino in formation of the lightest chargino. If the low scale spectrum includes two charginos lighter than about a TeV, then the processes can contribute to muon $g-2$, even when the heaviest chargino runs in the loop. Even though the heaviest chargino contribution can only be minor in comparison to the lightest chargino contribution, its mass cannot be heavier than about 800 GeV for the resolution to muon $g-2$ discrepancy, as seen from the  $\Delta a_{\mu}-m_{\tilde{\chi}_{2}^{\pm}}$. Finally, we also present the mass ratio of $M_{1}$ and $\mu$ in the $\Delta a_{\mu}-M_{1}/\mu$ plane. Even though $M_{1}$ does not interfere  in the SUSY contributions to muon $g-2$ due to the heavy smuons, the results in this plane reveal nature of LSP neutralino. Since $|M_{1}/\mu| \lesssim 1$, the Bino mixes in the LSP neutralino formation more than the Higgsinos. Comparing this plane with the results shown in Figure \ref{fig3}, one can easily see $M_{2} \lesssim M_{1}$, hence the model yields Wino-like LSP neutralino at the low scale. In addition, many of the solutions yield significant mixture of the neutralinos.

The light Higgsinos are also interesting from the naturalness point of view. Since the mass bounds on the supersymmetric particles become severe after the latest results from the LHC experiments, the solutions can barely be placed in the natural region characterized with $m_{\tilde{t}_{1}},m_{\tilde{t}_{2}}, m_{\tilde{b}_{1}} \lesssim 500$ GeV. Especially the Higgs boson mass constraint requires at least one stop to have mass above TeV scale. On the other hand, deviation from the natural region can be measured with $\Delta_{EW}$, the fine-tuning parameter as defined in Ref. \cite{Baer:2012mv}. $\Delta_{EW}$ is a function of $\mu$, $m_{H_{d}}$, $m_{H_{u}}$, and $\tan\beta$, in principal; however, the terms proportional to $m_{H_{d}}$ are suppressed by $\tan\beta$, and the correct electroweak symmetry breaking scale requires $\mu \approx m_{H_{u}}$ over most of the fundamental parameter space. In this sense, the Higgsino masses can also indicate the fine-tuning amount required to have consistent electroweak symmetry breaking. Since our model predicts relatively light Higgsinos ($\lesssim 800$ GeV) compatible with the resolution to muon $g-2$ discrepancy, such solutions also need low fine-tuning. In general fashion, the acceptable fine-tuning is identified with the condition $\Delta_{EW} \leq 1000$. Figure \ref{fig6} investigates our discussion about the fine-tuning with plots in the $\Delta a_{\mu}-\Delta_{EW}$, $y_{t}/y_{b}-\Delta_{EW}$, $y_{\tau}/y_{b}-\Delta_{EW}$ and $y_{\nu}/y_{b}-\Delta_{EW}$ planes. The color coding is the same as Figure \ref{fig3}. As seen from the  $\Delta a_{\mu}-\Delta_{EW}$ plane, $\Delta_{EW}$ can be as low as 20 compatible with muon $g-2$ condition. Indeed, muon $g-2$ condition restricts $\Delta_{EW} \lesssim 100$, and the discrepancy cannot be solved within $1\sigma$ when $\Delta_{EW} > 100$. In addition to muon $g-2$ resolution, we also discuss the Yukawa couplings, whose deviations are also restricted by the QYU condition in the $y_{t}/y_{b}-\Delta_{EW}$, $y_{\tau}/y_{b}-\Delta_{EW}$ and $y_{\nu}/y_{b}-\Delta_{EW}$ planes. According to the results represented in these planes, $y_{t}/y_{b} \gtrsim 2$, $y_{\tau}/y_{b}\gtrsim 1.2$, and $y_{\nu}/y_{b} \gtrsim 0.8$ over the region with the acceptable fine-tuning.

\section{Conclusion}
\label{sec:conc}

We explore the low scale implications of $4-2-2$ including the TeV scale right-handed neutrinos interacting and mixing with the MSSM fields through the IS mechanism, in light of muon $g-2$ resolution and highlight the solutions which are compatible with the QYU condition. We found that the presence of the right-handed neutrinos cause the smuons are rather heavy as $m_{\tilde{\mu}} \gtrsim 800$ GeV in order to avoid tachyonic staus at the low scale. In this context, the usual MSSM contributions to muon $g-2$, which are provided from smuon-neutralino loop, is strongly suppressed. On the other hand, the sneutrinos can be as light as about 100 GeV along with the light charginos of mass $\lesssim 400$ GeV can yield so large contributions to muon $g-2$ that the discrepancy between the experiment and the theory can be resolved. In addition, the model predicts relatively light Higgsinos ($\mu \lesssim 700$ GeV); and hence the second chargino mass is also light enough ($\lesssim 700$ GeV) to contribute to muon $g-2$. Despite the Higgsino mixing in the lightest neutralino and chargino is limited, the light Higgsinos are interesting from the naturalness point of view, since such solutions of the light Higgsinos need to be fine-tuned much less than the other solutions. We found that such solutions can be also compatible with the QYU, since $\Delta_{EW}$ can be as low as about 100. The acceptable fine-tuning can also have a strong impact on the Yukawa couplings in terms of their ratios, and this impact also shapes the fundamental parameter space of QYU, since it is rather related to the corrections in the Yukawa couplings. In the regions with acceptable fine-tuning and compatible with muon $g-2$ resolution and the QYU condition, the ratios among the Yukawa couplings can be summarized as $1.8 \lesssim y_{t}/y_{b} \lesssim 2.6$, $y_{\tau}/y_{b}\sim 1.3 $. In addition, even though the right-handed neutrino Yukawa coupling can be varied freely, the solutions restrict its range as $0.8\lesssim y_{\nu}/y_{b} \lesssim 1.7$.

\vspace{0.3cm}
\noindent {\bf Acknowledgments}

We would like to thank Zerrin K\i rca, B\"{u}\c{s}ra Ni\c{s} and Ali \c{C}i\c{c}i for discussions and complementary suggestions. This work is supported by the Scientific and Technological Research Council of Turkey (TUBITAK) Grant no. MFAG-114F461. Part of numerical calculations reported in this paper were performed at the National Academic Network and Information Center (ULAKBIM) of TUBITAK, High Performance anf Grid Computing Center (TRUBA Resources).


\begin{thebibliography}{99}

\bibitem{Wendell:2010md} 
  R.~Wendell {\it et al.} [Super-Kamiokande Collaboration],
  Phys.\ Rev.\ D {\bf 81}, 092004 (2010)
  doi:10.1103/PhysRevD.81.092004
  [arXiv:1002.3471 [hep-ex]].
  
  
\bibitem{big-422}
B. Ananthanarayan, G. Lazarides and Q. Shafi, Phys. Rev. D {\bf 44},
1613 (1991; Phys. Lett. B {\bf 300}, 24 (1993)5; Q.~Shafi and
B.~Ananthanarayan, Trieste HEP Cosmol.1991:233-244;
J.~C.~Pati and A.~Salam,
  Phys.\ Rev.\  D {\bf 10}, 275 (1974).  
  
\bibitem{pati-salam}
J.~C.~Pati and A.~Salam,
  Phys.\ Rev.\  D {\bf 10}, 275 (1974).
  
  
  
\bibitem{bigger-422}See, incomplete list of references,
L.~J.~Hall, R.~Rattazzi and U.~Sarid, Phys.\ Rev.\  D {\bf 50}, 7048 (1994);
 B. Ananthanarayan, Q. Shafi and X.
Wang, Phys. Rev. D {\bf 50}, 5980 (1994);
R. Rattazzi and U. Sarid, Phys. Rev. D {\bf 53}, 1553
(1996); T. Blazek, M. Carena, S. Raby and C. Wagner, Phys. Rev. D
{\bf 56}, 6919 (1997);
J.~L.~Chkareuli and I.~G.~Gogoladze,
  Phys.\ Rev.\  D {\bf 58}, 055011 (1998);
T. Blazek, S. Raby and K. Tobe, Phys. Rev. D
{\bf 62}, 055001 (2000); H. Baer, M. Brhlik, M. Diaz,
J. Ferrandis, P. Mercadante, P. Quintana and X. Tata, Phys. Rev. D
{\bf 63}, 015007(2001);  C.~Balazs and R.~Dermisek, JHEP {\bf 0306}, 024 (2003);
 U. Chattopadhyay, A. Corsetti and P.
Nath, Phys. Rev. D {\bf 66} 035003, (2002); T.~Blazek, R.~Dermisek
and S.~Raby, Phys.\ Rev.\ Lett.\  {\bf 88}, 111804 (2002); M. Gomez, T. Ibrahim, P. Nath and
S. Skadhauge, Phys. Rev. D {\bf 72}, 095008 (2005); K. Tobe and J.
D. Wells, Nucl. Phys. B {\bf 663}, 123 (2003);
I.~Gogoladze, Y.~Mimura, S.~Nandi,
  Phys.\ Lett.\  {\bf B562}, 307 (2003);
W.~Altmannshofer,
D.~Guadagnoli, S.~Raby and D.~M.~Straub, Phys.\ Lett.\  B {\bf 668},
385 (2008);
S.~Antusch and M.~Spinrath,
 Phys.\ Rev.\  D {\bf 78}, 075020 (2008);
 H.~Baer, S.~Kraml and S.~Sekmen, JHEP {\bf 0909}, 005 (2009);
S.~Antusch and M.~Spinrath,
Phys.\ Rev.\  D {\bf 79}, 095004 (2009);
K.~Choi, D.~Guadagnoli, S.~H.~Im and C.~B.~Park,
  JHEP {\bf 1010}, 025 (2010);
  M.~Badziak, M.~Olechowski and S.~Pokorski,
  JHEP\ {\bf 1108}, 147  (2011);
  S.~Antusch, L.~Calibbi, V.~Maurer, M.~Monaco and M.~Spinrath,
   Phys.\ Rev.\ D {\bf 85}, 035025 (2012).
  J.~S.~Gainer, R.~Huo and C.~E.~M.~Wagner,
    JHEP {\bf 1203}, 097 (2012);
H.~Baer, S.~Raza and Q.~Shafi,
  Phys.\ Lett.\ B {\bf 712}, 250 (2012);
  I.~Gogoladze, Q.~Shafi, C.~S.~Un and ,
  JHEP {\bf 1207}, 055 (2012);
  M.~Badziak,
  Mod.\ Phys.\ Lett.\ A {\bf 27}, 1230020 (2012);
  G.~Elor, L.~J.~Hall, D.~Pinner and  J.~T.~Ruderman,
  JHEP {\bf 1210}, 111 (2012).
  I.~Gogoladze, Q.~Shafi and C.~S.~Un,
  Phys.\ Lett.\ B {\bf 704}, 201 (2011)
  I.~Gogoladze, Q.~Shafi and C.~S.~Un,
  JHEP {\bf 1208}, 028 (2012)
  M.~A.~Ajaib, I.~Gogoladze and Q.~Shafi,
  arXiv:1307.4882 [hep-ph].
  M.~Adeel Ajaib, I.~Gogoladze, Q.~Shafi and C.~S.~Un,
  JHEP {\bf 1307}, 139 (2013)
  M.~A.~Ajaib, I.~Gogoladze, Q.~Shafi and C.~S.~Un,
  arXiv:1308.4652 [hep-ph].
  
\bibitem{Gogoladze:2010fu} 
  I.~Gogoladze, R.~Khalid, S.~Raza and Q.~Shafi,
  JHEP {\bf 1012}, 055 (2010);
%
  D.~M.~Pierce, J.~A.~Bagger, K.~T.~Matchev and R.~j.~Zhang,
  Nucl.\ Phys.\ B {\bf 491}, 3 (1997).

\bibitem{Langacker:1980js} 
  P.~Langacker,
  Phys.\ Rept.\  {\bf 72}, 185 (1981).

\bibitem{Witten:1979nr}
See, for instance, E.~Witten,
Phys.\ Lett.\  {\bf B91}, 81 (1980);
  S.~M.~Barr,
  Phys.\ Rev.\ D {\bf 21}, 1424 (1980);
  Y.~Nomura and T.~Yanagida,
  Phys.\ Rev.\ D {\bf 59}, 017303 (1999);
  M.~Frigerio, P.~Hosteins, S.~Lavignac and A.~Romanino,
  Nucl.\ Phys.\ B {\bf 806}, 84 (2009);
  S.~M.~Barr,
  Phys.\ Rev.\ D {\bf 76}, 105024 (2007);
  M.~Malinsky,
  Phys.\ Rev.\ D {\bf 77}, 055016 (2008);
  M.~Heinze and M.~Malinsky,
  Phys.\ Rev.\ D {\bf 83}, 035018 (2011);
  K.~S.~Babu, B.~Bajc and Z.~Tavartkiladze,
  Phys.\ Rev.\ D {\bf 86}, 075005 (2012)
   and references therein.

\bibitem{Babu:1992ia}
  K.~S.~Babu and R.~N.~Mohapatra,
  Phys.\ Rev.\ Lett.\  {\bf 70}, 2845 (1993).

\bibitem{Joshipura:2012sr}
 For a brief review, see  A.~S.~Joshipura and K.~M.~Patel,
  Phys.\ Rev.\ D {\bf 86}, 035019 (2012) and references therein.

\bibitem{Gomez:2002tj}
  M.~E.~Gomez, G.~Lazarides and C.~Pallis,
  Nucl.\ Phys.\ B {\bf 638}, 165 (2002)
  [hep-ph/0203131];
  M.~E.~Gomez, G.~Lazarides and C.~Pallis,
  Phys.\ Rev.\ D {\bf 67}, 097701 (2003)
  [hep-ph/0301064];
  C.~Pallis and M.~E.~Gomez,
  hep-ph/0303098;

\bibitem{Gogoladze:2009ug} 
  I.~Gogoladze, R.~Khalid and Q.~Shafi,
  Phys.\ Rev.\ D {\bf 79}, 115004 (2009)
  doi:10.1103/PhysRevD.79.115004
  [arXiv:0903.5204 [hep-ph]].

\bibitem{Raza:2014upa} 
  S.~Raza, Q.~Shafi and C.~S.~Ün,
  Phys.\ Rev.\ D {\bf 92}, no. 5, 055010 (2015)
  doi:10.1103/PhysRevD.92.055010
  [arXiv:1412.7672 [hep-ph]].

\bibitem{Dar:2011sj} 
  S.~Dar, I.~Gogoladze, Q.~Shafi and C.~S.~Un,
  Phys.\ Rev.\ D {\bf 84}, 085015 (2011);
  Q.~Shafi, Ş.~H.~Tanyıldızı and C.~S.~Un,
  Nucl.\ Phys.\ B {\bf 900}, 400 (2015);
  Y.~Hiçyılmaz, M.~Ceylan, A.~Altas, L.~Solmaz and C.~S.~Un,
  Phys.\ Rev.\ D {\bf 94}, no. 9, 095001 (2016).


\bibitem{Bajc:2004xe}
  B.~Bajc, A.~Melfo, G.~Senjanovic and F.~Vissani,
  Phys.\ Rev.\  D {\bf 70}, 035007 (2004);
  S.~Bertolini, M.~Frigerio, M.~Malinsky,
  Phys.\ Rev.\  {\bf D70}, 095002 (2004);
  B.~Dutta, Y.~Mimura, R.~N.~Mohapatra,
  Phys.\ Rev.\  {\bf D69}, 115014 (2004);
  T.~Fukuyama, A.~Ilakovac, T.~Kikuchi, S.~Meljanac and N.~Okada,
  J.\ Math.\ Phys.\  {\bf 46}, 033505 (2005).

\bibitem{Antusch:2013rxa} 
  S.~Antusch, S.~F.~King and M.~Spinrath,
  Phys.\ Rev.\ D {\bf 89}, no. 5, 055027 (2014);
  S.~Antusch, L.~Calibbi, V.~Maurer, M.~Monaco and M.~Spinrath,
  Phys.\ Rev.\ D {\bf 85}, 035025 (2012);
  S.~Antusch and M.~Spinrath,
  Phys.\ Rev.\ D {\bf 79}, 095004 (2009);
  S.~Trine, S.~Westhoff and S.~Wiesenfeldt,
  JHEP {\bf 0908}, 002 (2009).
\bibitem{Hebbar:2017olk} 
  A.~Hebbar, Q.~Shafi and C.~S.~Un,
  arXiv:1702.05431 [hep-ph], and references therein.


\bibitem{Coriano:2014wxa} 
  C.~Coriano, L.~Delle Rose and C.~Marzo,
  Nucl.\ Part.\ Phys.\ Proc.\  {\bf 265-266}, 311 (2015);
  S.~Khalil and H.~Okada,
  Prog.\ Theor.\ Phys.\ Suppl.\  {\bf 180}, 35 (2010);
  M.~Abbas and S.~Khalil,
  JHEP {\bf 0804}, 056 (2008), and references therein.

\bibitem{Khalil:2010iu} 
  S.~Khalil,
  Phys.\ Rev.\ D {\bf 82}, 077702 (2010);
  A.~Masiero, S.~K.~Vempati and O.~Vives,
  Nucl.\ Phys.\ Proc.\ Suppl.\  {\bf 137}, 156 (2004).

\bibitem{Bennett:2006fi} 
  G.~W.~Bennett {\it et al.} [Muon g-2 Collaboration],
  Phys.\ Rev.\ D {\bf 73}, 072003 (2006);
  G.~W.~Bennett {\it et al.} [Muon (g-2) Collaboration],
  Phys.\ Rev.\ D {\bf 80}, 052008 (2009).

\bibitem{Davier:2010nc}
  M.~Davier, A.~Hoecker, B.~Malaescu and Z.~Zhang,
  Eur.\ Phys.\ J.\ C {\bf 71}, 1515 (2011)
  [Eur.\ Phys.\ J.\ C {\bf 72}, 1874 (2012)]
  [arXiv:1010.4180 [hep-ph]];
  K.~Hagiwara, R.~Liao, A.~D.~Martin, D.~Nomura and T.~Teubner,
  J.\ Phys.\ G {\bf 38}, 085003 (2011)
  [arXiv:1105.3149 [hep-ph]].


\bibitem{Moroi:1995yh}
  T.~Moroi,
  Phys.\ Rev.\ D {\bf 53}, 6565 (1996)
  [Phys.\ Rev.\ D {\bf 56}, 4424 (1997)]
  [hep-ph/9512396];
  S.~P.~Martin and J.~D.~Wells,
  Phys.\ Rev.\ D {\bf 64}, 035003 (2001)
  [hep-ph/0103067];
  G.~F.~Giudice, P.~Paradisi, A.~Strumia and A.~Strumia,
  JHEP {\bf 1210}, 186 (2012)
  [arXiv:1207.6393 [hep-ph]].

\bibitem{Khalil:2015wua} 
  S.~Khalil and C.~S.~Un,
  Phys.\ Lett.\ B {\bf 763}, 164 (2016)

\bibitem{Gogoladze:2014vea} 
  I.~Gogoladze, B.~He, A.~Mustafayev, S.~Raza and Q.~Shafi,
  JHEP {\bf 1405}, 078 (2014). 

\bibitem{Babu:2014lwa} 
  K.~S.~Babu, I.~Gogoladze, Q.~Shafi and C.~S.~Ün,
  Phys.\ Rev.\ D {\bf 90}, no. 11, 116002 (2014);
  M.~A.~Ajaib, I.~Gogoladze, Q.~Shafi and C.~S.~Ün,
  JHEP {\bf 1405}, 079 (2014);
  I.~Gogoladze, F.~Nasir, Q.~Shafi and C.~S.~Un,
  Phys.\ Rev.\ D {\bf 90}, no. 3, 035008 (2014);
  I.~Gogoladze, Q.~Shafi and C.~S.~Ün,
  Phys.\ Rev.\ D {\bf 92}, no. 11, 115014 (2015);
  B.~P.~Padley, K.~Sinha and K.~Wang,
  Phys.\ Rev.\ D {\bf 92}, no. 5, 055025 (2015); and references therein.



\bibitem{Baer:2012mv} See for instance, \\
  H.~Baer, V.~Barger, P.~Huang, D.~Mickelson, A.~Mustafayev and X.~Tata,
  Phys.\ Rev.\ D {\bf 87}, no. 3, 035017 (2013);
  D.~A.~Demir and C.~S.~Ün,
  Phys.\ Rev.\ D {\bf 90}, 095015 (2014);
  A.~Cici, Z.~Kirca and C.~S.~Un,
  arXiv:1611.05270 [hep-ph]; and references therein.

\bibitem{Porod:2003um}
  Porod, W.
  {\it Comput.\ Phys.\ Commun.\ } {\bf 2003}, {\it 153}, 275;  
  Porod, W. and Staub, F.
  {\it Comput.\ Phys.\ Commun.\ } {\bf 2012} {\it 183}, 2458.

\bibitem{Staub:2008uz}
  Staub, F.
{\bf 2008}, {\it Preprint  arXiv:0806.0538}.
  
  \bibitem{Hisano:1992jj}
  Hisano, J.; Murayama, H.; and Yanagida, T.
 {\it Nucl.\ Phys.\ B } {\bf 1993} {\it 402}, 46;
  Chkareuli, J. L.; and Gogoladze, I. G.
 {\it Phys.\ Rev.\ D }{\bf 1998} {\it 58}, 055011.
  
  \bibitem{Group:2009ad}
  T.~E.~W.~Group [CDF and D0 Collaborations], {\bf 2009}, {\it Preprint
  arXiv:0903.2503}.
  
  \bibitem{Gogoladze:2011db}
  Gogoladze, I.; Khalid, R.; Raza S.; and Shafi Q.
  {\it JHEP } {\bf 2011}, {\it 1106}, 117.
\bibitem{Gogoladze:2011aa}
  Gogoladze, I.; Shafi, Q.; and Un, C. S.
 {\it JHEP} {\bf 2012} {\it 1208}, 028;
  Adeel Ajaib, M.; Gogoladze, I.; Shafi, Q.; and Un, C. S.
  {\it JHEP } {\bf 2013} {\it 1307}, 139.
  
\bibitem{Ibanez:Ross}
  Ibanez, L. E.; and Ross, G. G.
  {\it Phys.\ Lett.} {\it 110B} {\bf 1982} 215;
  Inoue, K.; Kakuto, A.; Komatsu, H.; and Takeshita S.,
  {\it Prog.\ Theor.\ Phys.\ }  {\bf 1982} {\it 68}, 927;
  Ibanez, L. E.
  {\it Phys.\ Lett.\ } {\bf 1982} {\it 118B}, 73;
  Ellis, J. R.; Nanopoulos D. V.; and Tamvakis, K.
 {\it Phys.\ Lett.\ } {\bf 1983} {\it 121B}, 123;
   Alvarez-Gaume, L.; Polchinski, J.; and Wise, M. B.
 {\it Nucl.\ Phys.\ B} {\bf 1983} {\it 221}, 495.  
 
\bibitem{Nakamura:2010zzi}
  Nakamura, K. {\it et al.} [Particle Data Group Collaboration],
  {\it J.\ Phys.\ G} {\bf 2010} {\it 37}, 075021.
  
\bibitem{Baer:2012by} See for instance;

  H.~Baer, I.~Gogoladze, A.~Mustafayev, S.~Raza and Q.~Shafi,
  JHEP {\bf 1203}, 047 (2012)
  doi:10.1007/JHEP03(2012)047
  [arXiv:1201.4412 [hep-ph]];
  T.~Li, D.~V.~Nanopoulos, S.~Raza and X.~C.~Wang,
  JHEP {\bf 1408}, 128 (2014)
  doi:10.1007/JHEP08(2014)128
  [arXiv:1406.5574 [hep-ph]].
  
  
  
\bibitem{Belanger:2009ti}
  Belanger, G.; Boudjema, F.; Pukhov, A.; and Singh, R. K.
  {\it JHEP } {\bf 2009} {\it 0911}, 026;
  Baer, H.; Kraml, S.; Sekmen, S.; and Summy, H.
  {\it JHEP } {\bf 2008} {\it 0803}, 056.

\bibitem{Agashe:2014kda}
  Olive, K. A. {\it et al.} [Particle Data Group Collaboration],
  {\it Chin.\ Phys.\ C } {\bf 2014} {\it 38}, 090001.

\bibitem{Aaij:2012nna}
  Aaij, R. {\it et al.} [LHCb Collaboration],
  {\it Phys.\ Rev.\ Lett.\ } {\bf 2013}  {\it 110}, no. 2, 021801.

\bibitem{Amhis:2012bh} 
  Y.~Amhis {\it et al.} [Heavy Flavor Averaging Group Collaboration],
  arXiv:1207.1158 [hep-ex].

\bibitem{Aad:2012tfa} 
  Aad, G.; {\it et al.} [ATLAS Collaboration],
  {\it Phys.\ Lett.\ B} {\bf 2012}, {\it 716}, 1-29; 
 Chatrchyan, S.; {\it et al.} [CMS Collaboration],
  {\it JHEP} {\bf 2013}, {\it 1306}, 081.
  
  \bibitem{TheATLAScollaboration:2015aaa}
  The ATLAS collaboration,
 {\it  ATLAS-CONF-2015-067} {\bf 2015}.

\bibitem{Gogoladze:2009bd} 
  I.~Gogoladze, M.~U.~Rehman and Q.~Shafi,
  Phys.\ Rev.\ D {\bf 80}, 105002 (2009)
  doi:10.1103/PhysRevD.80.105002
  [arXiv:0907.0728 [hep-ph]].

\bibitem{Antusch:2011xz} 
  S.~Antusch, L.~Calibbi, V.~Maurer, M.~Monaco and M.~Spinrath,
  Phys.\ Rev.\ D {\bf 85}, 035025 (2012)
  doi:10.1103/PhysRevD.85.035025
  [arXiv:1111.6547 [hep-ph]].



\end{thebibliography}
\end{document}